\documentclass{article}
\usepackage{arxiv}

\usepackage[utf8]{inputenc} 
\usepackage[T1]{fontenc}    
\usepackage{hyperref}       
\usepackage{url}            
\usepackage{booktabs}       
\usepackage{amsfonts}       
\usepackage{nicefrac}       
\usepackage{microtype}      
\usepackage{lipsum}		
\usepackage{graphicx}
\usepackage[square,numbers]{natbib}
\usepackage{doi}
\usepackage{placeins} 
\usepackage{xcolor}
\usepackage{multirow}
\usepackage{multicol}
\usepackage{amsmath}
\usepackage{array}
\usepackage{xspace} 
\newcolumntype{C}[1]{>{\centering\arraybackslash}p{#1}}

\title{Lattice-to-Total Thermal Conductivity Ratio: A Phonon-Glass Electron-Crystal Descriptor for Data-Driven Thermoelectric Design} 

\author{ \hspace{1mm}Yifan Sun \thanks{These authors contributed equally to this work.}  \\
	Kyoto University\\
	\And
	\hspace{1mm}Zhi Li \footnotemark[1]\\
	Northwestern University\\
    \And
	\hspace{1mm}Tetsuya Imamura\\
	Osaka University\\
	\And
	\hspace{1mm}Yuji Ohishi\\
	Osaka University\\
	\And
	\hspace{1mm}Chris Wolverton \thanks{Correspondence to: c-wolverton@northwestern.edu}\\
	Northwestern University\\
    \And
	\hspace{1mm}Ken Kurosaki \thanks{Correspondence to: kurosaki.ken.6n@kyoto-u.ac.jp} \\
	Kyoto University\\
}

\date{}

\renewcommand{\headeright}{}
\renewcommand{\undertitle}{}
\renewcommand{\shorttitle}{}
\newcommand{\wmk}[0]{$~\mathrm{W\,m^{-1}\,K^{-1}}$\xspace}
\newcommand{\klk}[0]{$\kappa_\mathrm{L}/\kappa$\xspace}

\begin{document}
\maketitle

\begin{abstract}
Thermoelectrics (TEs) are promising candidates for energy harvesting with performance quantified by figure of merit, $ZT$. To accelerate the discovery of high-$ZT$ materials, efforts have focused on identifying compounds with low thermal conductivity $\kappa$. Using a curated dataset of 71{,}913 entries, we show that high-$ZT$ materials reside not only in the low-$\kappa$ regime but also cluster near a lattice-to-total thermal conductivity ratio (\klk) of approximately 0.5. This optimal ratio provides a quantitative descriptor for the well-known phonon-glass electron-crystal (PGEC) design concept. Building on this insight, we construct a framework consisting of two machine learning models for the lattice and electronic components of thermal conductivity that jointly provide both $\kappa$ and \klk for screening and guiding the optimization of TE materials. By applying this framework to 104{,}567 inorganic compounds, we identify 2{,}522 ultralow-$\kappa$ candidates while simultaneously evaluating their proximity to the optimal PGEC regime. A follow-up case study on chemical doping demonstrates how the framework can qualitatively provide optimization strategies that shift pristine materials toward the ideal $\kappa_\mathrm{L}/\kappa$ $\approx$ 0.5 target. Ultimately, by integrating rapid screening with PGEC-guided optimization, our data-driven framework takes a critical step towards closing the gap between materials discovery and performance enhancement.
\end{abstract}



\section{Introduction}
More than half of global primary energy consumption is ultimately lost after conversion, much of it as low- to medium-grade heat from industrial processes, transportation, and power generation~\cite{forman2016estimating,geffroy2021techno}. Recovering even a modest portion of this waste heat could significantly improve overall energy efficiency and contribute to sustainability goals. Thermoelectric (TE) materials, which can directly convert heat into electricity, are therefore attractive candidates for waste-heat harvesting in applications ranging from automotive systems to industrial plants~\cite{bell2008cooling,snyder2008complex}. The conversion efficiency of TE materials is commonly evaluated using the dimensionless figure of merit $ZT$, defined as
\begin{equation}
\label{eqn.zt}
    ZT = \frac{S^2 \sigma}{\kappa_\mathrm{L} + \kappa_\mathrm{e}}\, T,
\end{equation}
where \(S\) is the Seebeck coefficient, \(\sigma\) is the electrical conductivity, $T$ is the absolute temperature, and \(\kappa_\mathrm{L}\) and \(\kappa_\mathrm{e}\) are the lattice and carrier contributions to the total thermal conductivity $\kappa$. A high $ZT$ requires a large power factor $PF = S^2\sigma$ and a low $\kappa$. However, optimizing TE performance is challenging because these transport parameters are strongly interrelated, and improving one often comes at the expense of another. These intricate relationships make it difficult to identify high-$ZT$ materials through conventional trial-and-error experiments, and have motivated data-driven strategies, particularly machine learning (ML), to guide the search and design of TE materials more efficiently.

In recent years, while several ML models have been developed to predict the $ZT$ of materials directly~\cite{na2022public,li2022large,parse2023machine,barua2024thermoelectric,xu2024prediction}, direct prediction often lacks physical interpretability because $ZT$ is a composite metric governed by competing mechanisms. Consequently, much of the ML effort has shifted toward predicting individual TE properties. Models have been trained to predict quantities such as the Seebeck coefficient $S$~\cite{na2022public,furmanchuk2018prediction,yuan2022machine,barua2025machine} and the thermal conductivity $\kappa$, especially its lattice component $\kappa_\mathrm{L}$~\cite{na2022public,tewari2020machine,wang2020identification,barua2024interpretable,li2025machine}. In particular, an intrinsically low $\kappa_\mathrm{L}$ is often used as a key indicator of a promising TE material, since $\kappa_\mathrm{L}$ is largely dictated by crystal structure and bonding and is less dependent on extrinsic carrier concentration than $S$ or $\sigma$. For example, Wang et al.~\cite{wang2020identification} used computed $\kappa_\mathrm{L}$ values from AFLOW to train an XGBoost model and identified several low-$\kappa$ candidate materials that were subsequently validated by first-principles calculations. Barua et al.~\cite{barua2024interpretable} developed ML models to predict $\kappa$ using the Starrydata database of experimental thermoelectric measurements. More recently, Li et al.~\cite{li2025machine} trained a neural network on first-principles $\kappa_\mathrm{L}$ data and screened over 32,000 compounds from the Inorganic Crystal Structure Database (ICSD), discovering numerous additional low-$\kappa$ candidates.

Still, finding a material with intrinsically low thermal conductivity is only a first step toward achieving a high $ZT$. In practice, most TE materials require further optimization, such as defect engineering~\cite{mao2017defect,zhang2023defect}, band structure engineering~\cite{pei2017integrating,xie2022high}, and nanostructuring~\cite{zhang2024high,zheng2015mechanically}, in order to boost electrical transport while further suppressing lattice thermal transport. The accumulated lessons from decades of such optimization efforts are encapsulated in the phonon-glass electron-crystal (PGEC) paradigm~\cite{slack1995crc,nolas1999skutterudites,takabatake2014phonon,snyder2004disordered}, which calls for materials that conduct heat like a glass (suppressing phonon mobility) but conduct electrons like a crystal (enhancing carrier mobility). In essence, the PGEC concept emphasizes that an ideal thermoelectric should strike a balance between phonon and carrier transport properties. To date, however, the explicit quantification and incorporation of a PGEC-based metric within data-driven screening pipelines remains conspicuously absent. To address this gap and move beyond simply screening for pristine materials with low $\kappa$ or $\kappa_{L}$, we adopt a data-driven approach to explicitly quantify the PGEC concept.

We first curate a large experimental dataset of thermoelectric properties from the Starrydata~\cite{katsura2025starrydata} repository and calculate the lattice ($\kappa_\mathrm{L}$) and electronic thermal conductivity ($\kappa_\mathrm{e}$) for each entry in a consistent manner. Here, $\kappa_\mathrm{L}$ reflects phonon transport, while $\kappa_\mathrm{e}$ serves as a proxy for carrier transport. Utilizing $\kappa_{e}$ is particularly advantageous because it shares the same physical dimension as $\kappa_{L}$, allowing for a direct and mathematically intuitive formulation of the thermal transport balance. Statistical analysis reveals that existing high-$ZT$ TE materials not only lie in the low-$\kappa$ regime ($\kappa \leq 2$~\wmk) but, more importantly, have $\kappa_\mathrm{L}/\kappa$ values clustering near 0.5. Further investigation into two representative TE families, CoSb$_3$ and GeTe, shows that, after the initial screening with $\kappa$, $\kappa_\mathrm{L}/\kappa$ acts as a valuable descriptor for designing optimization strategies toward the PGEC ideal, where $\kappa_\mathrm{L}/\kappa \approx 0.5$.

Consequently, to incorporate this PGEC-based descriptor into the screening and optimization of thermoelectric materials, we train two ML models, one to predict $\kappa_\mathrm{L}$ and another to predict $\kappa_\mathrm{e}$ from a set of composition descriptors~\cite{ward2016general} and temperature. These models allow us to obtain both $\kappa$ and $\kappa_\mathrm{L}/\kappa$, capture more granular information on phonon versus carrier thermal transport. To demonstrate the practical utility of our approach, we apply the trained models to screen 104{,}567 inorganic compounds from the Materials Project database~\cite{jain2013commentary} and identify 2{,}522 thermodynamically stable semiconductors with low thermal conductivity ($\kappa \leq 2$~\wmk) at 300~K. Follow-up validations against recent literature lend credence to the models' generalizability. Beyond macroscopic screening, we use AgBiS$_{2}$ as a case study to illustrate how our models can pinpoint specific elemental dopants capable of tuning the host matrix closer to the PGEC optimum (\klk$\rightarrow 0.5$). Overall, our data-driven framework, built on $\kappa_\mathrm{L}$ and $\kappa_\mathrm{e}$ models, not only enables the rapid identification of TE materials with intrinsically low thermal conductivity but also provides data-driven suggestions for optimizing those materials guided by the PGEC descriptor $\kappa_\mathrm{L}/\kappa$, thereby helping to bridge the gap between materials discovery and performance enhancement.

\section{Results}
\label{sec:data-drive-analysis}
\subsection{Data overview}
Conventionally, thermal conductivity, particularly the lattice component $\kappa_\mathrm{L}$, has long served as a primary descriptor in the search for promising thermoelectric materials. Because most pristine TE compounds are non-degenerate semiconductors with intrinsically low $\kappa_\mathrm{e}$, a low $\kappa_\mathrm{L}$ (or equivalently, low $\kappa$) is traditionally regarded as an essential prerequisite of high thermoelectric potential. However, materials exhibiting intrinsically low $\kappa_\mathrm{L}$ only provide a promising starting point. The subsequent optimization pathway needed to approach the PGEC regime, which is crucial from an experimentalist’s point of view, is not captured in most existing data-driven studies.

The emergence of the Starrydata~\cite{katsura2025starrydata} repository offers an unprecedented opportunity to address this gap, as it aggregates hundreds of thousands of experimentally measured samples, including both pristine compounds and a wide range of doped and alloyed derivatives. This breadth allows, for the first time at scale, direct tracking of how key thermoelectric properties evolve under chemical modification and how these modifications influence $ZT$. Because interpreting the high-dimensional and strongly coupled relationships among $S$, $\sigma$, $\kappa$, and $ZT$ is challenging, we focus our data-driven analysis specifically on thermal conductivity and its components. This approach provides systematic insights into how chemical modifications steer materials toward or away from the thermoelectric optimum.

\begin{figure}[htbp]
    \centering
    \includegraphics[width=0.98\linewidth]{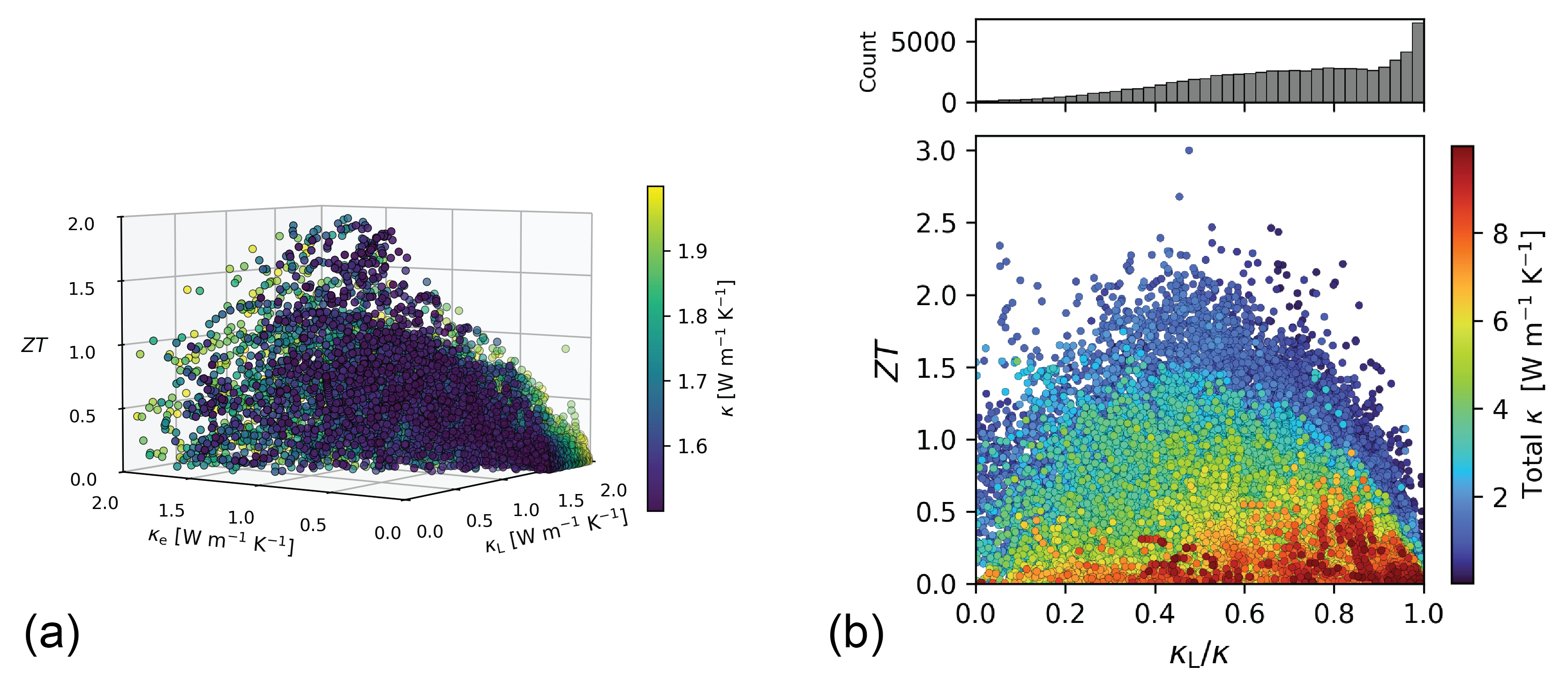}
    \caption{\textbf{Global relationships between $ZT$ and thermal conductivity distributions within the curated Starrydata dataset.} \textbf{a} 3D scatter plot illustrating the relation between $ZT$, lattice ($\kappa_{\mathrm{L}}$), and electronic ($\kappa_{\mathrm{e}}$) thermal conductivities at approximately fixed total $\kappa$. \textbf{b} $ZT$ as a function of the lattice-to-total thermal conductivity ratio ($\kappa_{\mathrm{L}}/\kappa$), revealing a distinct performance peak dynamically clustering around the phonon-glass electron-crystal (PGEC) optimum of $\sim$0.5.}
    \label{fig:data_driven_insight}
\end{figure}

First, we visualize the $ZT$ distribution with respect to $\kappa_\mathrm{e}$ and $\kappa_\mathrm{L}$ in Figure~\ref{fig:data_driven_insight}(a). For samples with approximately the same total thermal conductivity $\kappa$, the highest attainable $ZT$ tends to occur within a limited range of $\kappa_\mathrm{e}$ and $\kappa_\mathrm{L}$. To better clarify their relative contributions, we next plot $ZT$ with respect to the lattice-to-total thermal conductivity ratio $\kappa_\mathrm{L}/\kappa$ in Figure~\ref{fig:data_driven_insight}(b). Across the full dataset, $ZT$ varies with $\kappa_\mathrm{L}/\kappa$ in an inverted-U-shaped manner, with most data points lying in the $\kappa_\mathrm{L}/\kappa \approx 0.9$--1.0 range, while the highest $ZT$ values occur around $\kappa_\mathrm{L}/\kappa \approx 0.5$. From Figures~S4(a) and (b), we further confirm that the dominance of $\kappa_\mathrm{L}$ (with $\kappa_\mathrm{L}/\kappa$ close to 1) is present in both pristine and optimized samples, reflecting the typical behavior of non-degenerate semiconductors in which lattice thermal conductivity dominates. In contrast, the emergence of high $ZT$ near $\kappa_\mathrm{L}/\kappa \approx 0.5$ is evident only in the optimized samples, consistent with deliberate optimization efforts to enhance the carrier contribution while suppressing phonon transport. When we further divide the data by $\kappa$, as shown in Figure~S4(c), the characteristic $\kappa_\mathrm{L}/\kappa$ range associated with high $ZT$ shifts toward 0.5 as $\kappa$ decreases, indicating that the most successful optimizations drive materials toward a regime where lattice and carrier thermal transport contribute almost equally.

To illustrate more concretely how the tendency toward $\kappa_\mathrm{L}/\kappa \rightarrow 0.5$ reflects optimization toward the PGEC regime, we next examine the distribution of $ZT$ with respect to $\kappa_\mathrm{L}/\kappa$ before and after optimization in two representative families, CoSb$_3$ and GeTe (Figure~\ref{fig:zt_kl_k_family}). Details of the family classification are provided in the Supplementary Information.

\begin{figure}[htbp]
    \centering
    \includegraphics[width=0.48\linewidth]{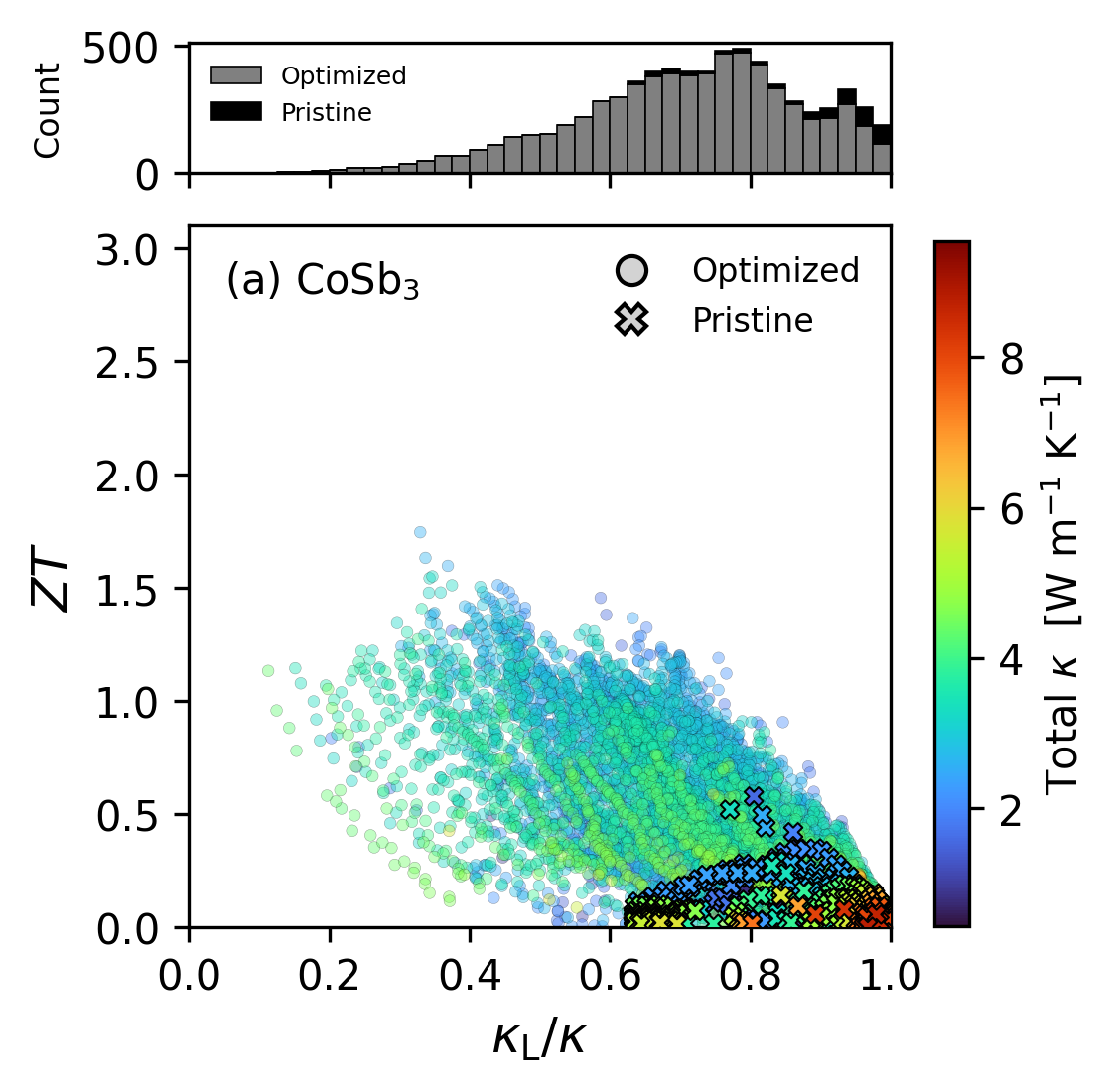}
    \includegraphics[width=0.48\linewidth]{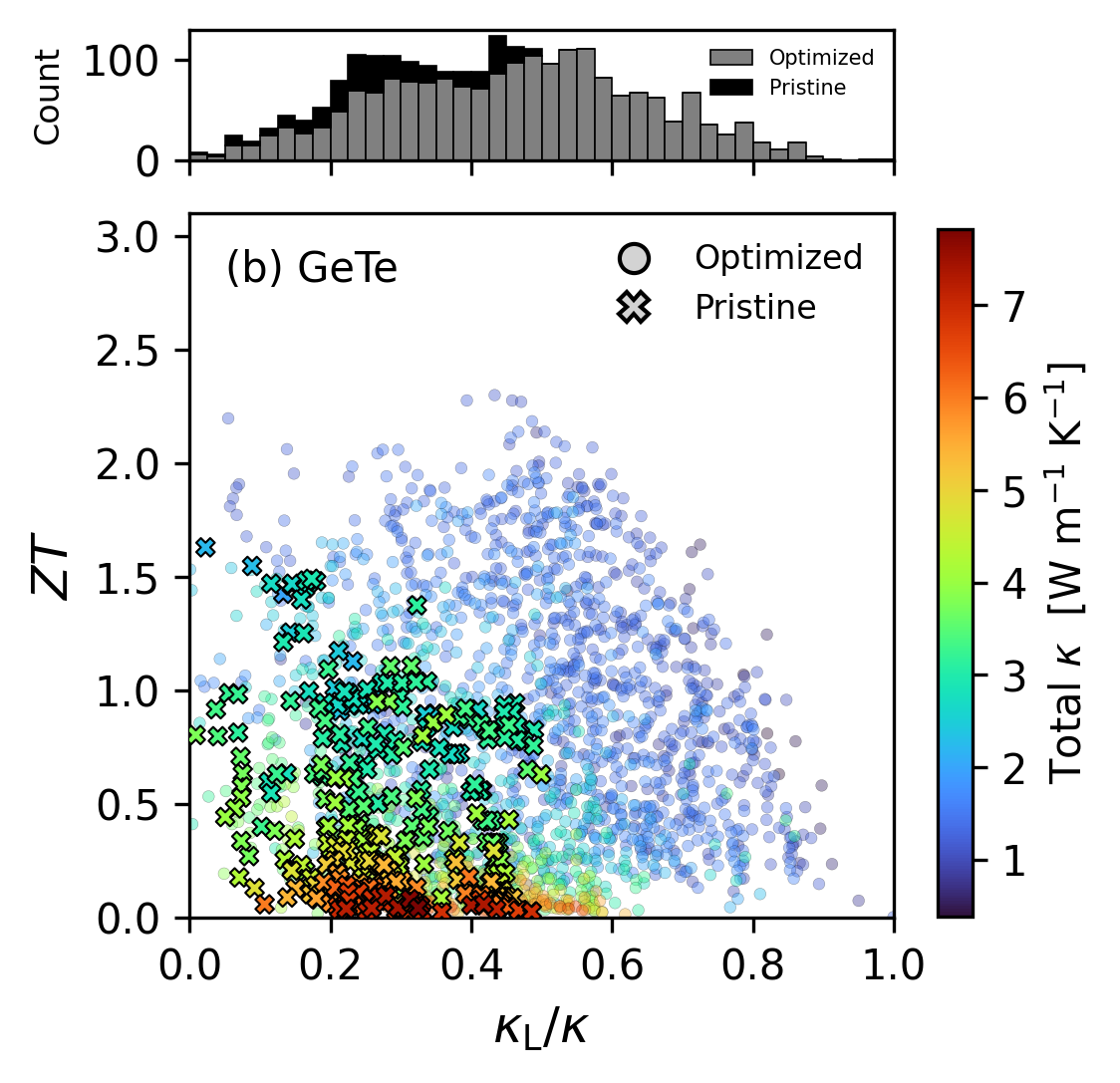}
    \caption{\textbf{Evolution of $ZT$ versus $\kappa_{\mathrm{L}}/\kappa$ during optimization for representative thermoelectric families.} \textbf{a} The intrinsically non-degenerate CoSb$_3$ family. \textbf{b} The inherently self-doped, degenerate GeTe family. Pristine compositions are denoted by filled crosses, whereas optimized variants are displayed as semi-transparent circles.}
    \label{fig:zt_kl_k_family}
\end{figure}

\paragraph{CoSb$_3$, a typical non-degenerate semiconductor.}
Figure~\ref{fig:zt_kl_k_family}(a) shows that unfilled skutterudites, CoSb$_3$, mostly have $ZT$ below 0.5 with $\kappa_\mathrm{L}/\kappa$ skewd towards the 0.6--1.0 range. This is because pristine CoSb$_3$ exhibits typical non-degenerate semiconductor behavior with low intrinsic carrier concentrations ($10^{17}$–$10^{18}$~cm$^{-3}$), resulting in a small $\kappa_\mathrm{e}$~\cite{lei2019ultrafast,shi2011multiple}. In addition, single-crystalline CoSb$_3$ has a relatively high intrinsic lattice thermal conductivity of $\kappa_\mathrm{L} \sim 10$~\wmk~\cite{caillat1996properties}. Consequently, the corresponding optimization strategy follows the classic thermoelectric playbook, reducing $\kappa_\mathrm{L}$ while maintaining or improving electrical transport,a dual approach that indirectly manifests as a stark reduction in $\kappa_\mathrm{L}/\kappa$. As a result, optimized CoSb$_3$ samples show reduced $\kappa$ and $\kappa_\mathrm{L}/\kappa$ values that move closer to 0.5.

Park et al.~\cite{park2014thermoelectric} provide a representative example of the outlined strategy by filling CoSb$_3$ with Yb. The intrinsic CoSb$_3$ sample (sample-id 4428) is a $p$-type semiconductor with a hole concentration of $2.56 \times 10^{18}$~cm$^{-3}$. Adding a small amount of Yb yields Yb$_{0.2}$Co$_4$Sb$_{12}$ (sample-id 4430), which becomes $n$-type with a much higher electron concentration of $2.45 \times 10^{20}$~cm$^{-3}$. At 623~K, this transformation changes $S$ from approximately +190 to $-200$~$\mu$V\,K$^{-1}$ and $\sigma$ from 208 to 1{,}024~S\,cm$^{-1}$, and is reflected in a jump in $\kappa_\mathrm{e}$ from 0.22 to 1.07~\wmk, confirming Yb as an efficient donor in CoSb$_3$. At the same time, weak bonding and large mass contrast make Yb act as a rattler in the CoSb$_3$ framework, reducing $\kappa_\mathrm{L}$ from 4.60 to 2.00~\wmk{} at 623~K. Together, these changes implement the PGEC strategy, raising the $ZT$ of CoSb$_3$ from 0.1 to 0.9 at 623~K and shifting $\kappa_\mathrm{L}/\kappa$ from 0.95 to 0.65, toward the $\kappa_\mathrm{L}/\kappa \approx 0.5$ PGEC regime.

\paragraph{GeTe, a self-doped degenerate semiconductor.}

In Figure~\ref{fig:zt_kl_k_family}(b), pristine GeTe samples (before optimization) generally show $\kappa_\mathrm{L}/\kappa$ values smaller than 0.5. This behavior can be traced to two main factors. First, strong anharmonicity induced by the stereochemically active lone pairs on Ge leads to a relatively low $\kappa_\mathrm{L}$~\cite{dou2021lone}. Second, pristine GeTe is well known for its very high intrinsic Ge-vacancy concentration~\cite{liu2018phase,li2017simultaneous,wu2017resonant}. These native acceptors boost the hole concentration to $\sim 10^{21}$~cm$^{-3}$, making GeTe an intrinsically degenerate semiconductor with high $\sigma$ and therefore a large $\kappa_\mathrm{e}$~\cite{levin2013electronic,hong2018realizing}. However, the excessive hole concentration also results in a low $S$, which explains why most pristine GeTe samples in Figure~\ref{fig:zt_kl_k_family}(b) still exhibit $ZT \leq 1$.

Upon optimization, we observe significantly increased $ZT$ accompanied by decreased $\kappa$ and an increase in $\kappa_\mathrm{L}/\kappa$ toward 0.5. A representative example is provided by Hong et al.~\cite{hong2018realizing}, who doped Sb on the Ge site. The Sb$^{3+}$ cations act as donors and reduce the hole concentration from $\sim10^{21}$~cm$^{-3}$ in pristine GeTe (sample-id 31967) to $\sim10^{20}$~cm$^{-3}$ in Ge$_{0.9}$Sb$_{0.1}$Te (sample-id 31969). The suppressed hole concentration increases $S$ and decreases $\sigma$, shifting $\kappa_\mathrm{L}/\kappa$ from 0.26 in pristine GeTe to 0.44 in the doped sample at 700~K, closer to the 0.5 threshold. At this temperature, pristine GeTe reaches its peak $ZT$ of 0.94, while the doped variant, benefiting from a reduced $\kappa$ and a more balanced phonon-carrier transport, nearly doubles that value to 1.68.

Taken together, the opposing optimization trajectories of CoSb$_3$ and GeTe reveal a profound insight: the optimal thermal transport conditions for high $ZT$ are remarkably universal across disparate chemical systems. Whether a material begins as a phonon-dominated host (\klk to be driven downwards) or an electron-flooded degenerate matrix (\klk to be driven upwards), high-performance samples consistently converge toward $\kappa_{\mathrm{L}}/\kappa \approx 0.5$ upon successful optimization. Thus, our data-driven analysis cements $\kappa_{\mathrm{L}}/\kappa$ not merely as an observation, but as an actionable, PGEC-based quantitative compass for guiding experimental optimization once an intrinsically low-$\kappa$ material has been successfully identified via high-throughput screening.

\subsection{Model Performances}

There are two critical reasons we decided to train separate $\kappa_\mathrm{L}$ and $\kappa_\mathrm{e}$ models rather than simply pairing $\kappa_\mathrm{L}$ with a single $\kappa$ model, which could technically also yield $\kappa_{\mathrm{L}}/\kappa$. First, $\kappa_\mathrm{L}$ and $\kappa_\mathrm{e}$ are governed by fundamentally different transport mechanisms. Training separate models allows each to learn the distinct compositional features relevant to its target property, as later demonstrated by the SHAP analysis. More importantly, $\kappa_\mathrm{e}$ is typically an order of magnitude smaller than $\kappa_\mathrm{L}$ in thermoelectric materials. Therefore, a model trained directly on $\kappa$ is expected to be dominated by the lattice component, effectively treating the $\kappa_\mathrm{e}$ variations as background noise. Consequently, paired $\kappa$ and $\kappa_\mathrm{L}$ models would learn similar features, rendering our PGEC descriptor less informative. By separately modeling $\kappa_\mathrm{e}$, we ensure that the electronic component is explicitly captured, preserving meaningful variation in $\kappa_{\mathrm{L}}/\kappa$ across different compositions.

The cross-validation results for both models ($\kappa_\mathrm{L}$ and $\kappa_\mathrm{e}$) averaged across five folds are summarized in Table~\ref{tab:model_cv_results}. For predicting both $\kappa_{\mathrm{L}}$ and $\kappa_{\mathrm{e}}$, we selected Random Forest as the best model based on the cross-validated mean absolute error (MAE). While CatBoost offers a slightly higher R$^2$ when predicting electronic thermal conductivity, the improvement is marginal (less than 0.01), and its much larger hyperparameter space makes tuning computational-extensive and increases the risk of overfitting on the training data.

\begin{table}[htbp]
    \centering
    \small
    \caption{\textbf{Cross-validation performance of candidate machine learning algorithms.} The table compares the predictive accuracy of four tree-based models for both lattice ($\kappa_{\mathrm{L}}$) and electronic ($\kappa_{\mathrm{e}}$) thermal conductivities. The Random Forest algorithm yields the lowest Mean Absolute Error (MAE) for both target properties (bolded). Units for MAE and Root Mean Squared Error (RMSE) are W\,m$^{-1}$\,K$^{-1}$.}
    \label{tab:model_cv_results}
    \begin{tabular}{lC{1.4cm}C{1.4cm}C{1.4cm}C{1.4cm}C{1.4cm}C{1.4cm}}
        \toprule
        \multirow{2}{*}{Model} &
        \multicolumn{3}{c}{Lattice thermal conductivity ($\kappa_{\mathrm{L}}$)} &
        \multicolumn{3}{c}{Electronic thermal conductivity ($\kappa_{\mathrm{e}}$)} \\
        \cmidrule(lr){2-4}\cmidrule(lr){5-7}
        & MAE & RMSE & R$^2$ & MAE & RMSE & R$^2$ \\
        \midrule
        XGBoost        & 0.38 & 0.66 & 0.77 & 0.25 & 0.41 & 0.71 \\
        LightGBM       & 0.39 & 0.65 & 0.78 & 0.26 & 0.41 & 0.71 \\
        Random Forest  & \textbf{0.35} & 0.64 & 0.79 & \textbf{0.23} & 0.39 & 0.73 \\
        CatBoost       & 0.37 & 0.63 & 0.79 & 0.24 & 0.39 & 0.74 \\
        \bottomrule
    \end{tabular}
\end{table}

\begin{figure}[htbp]
    \centering
    \includegraphics[width=0.48\linewidth]{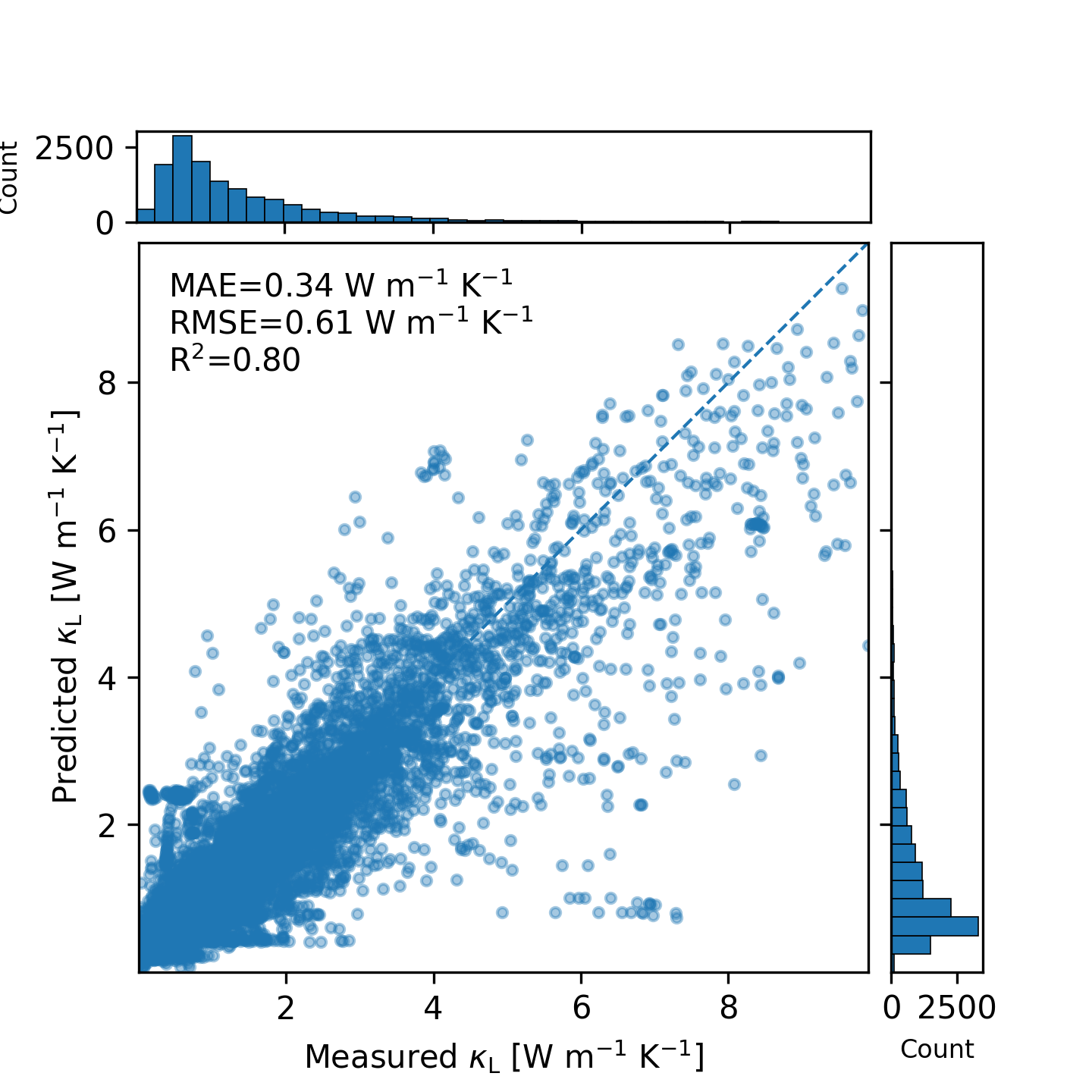}  
    \includegraphics[width=0.48\linewidth]{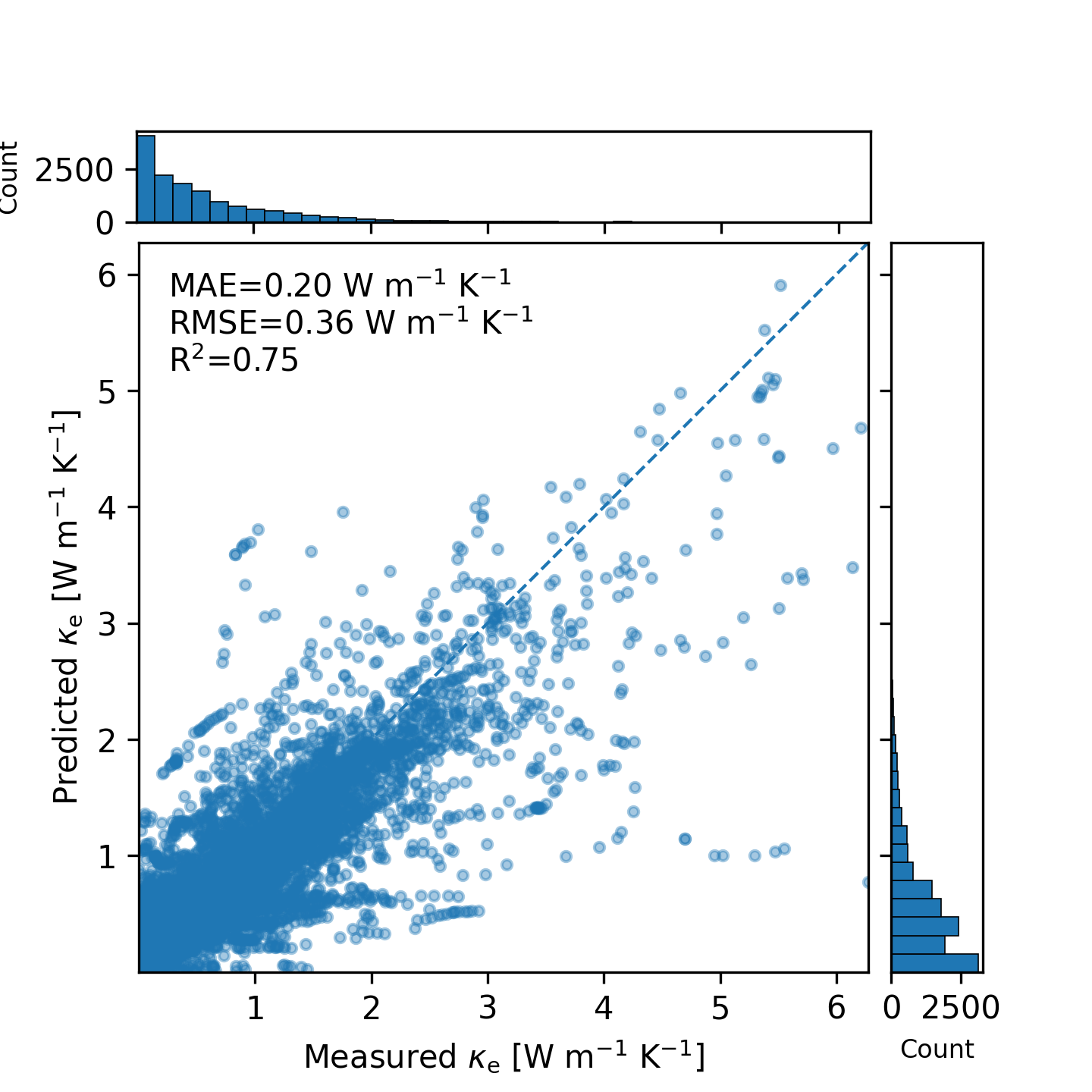}
    \caption{\textbf{Predictive performance of the fine-tuned Random Forest models on the held-out test set.} Parity plots comparing the machine learning predictions against experimental measurements for \textbf{a} lattice thermal conductivity ($\kappa_{\mathrm{L}}$) and \textbf{b} electronic thermal conductivity ($\kappa_{\mathrm{e}}$).}
    \label{fig:test_performances}
\end{figure}

After model selection, the search spaces we explored and the optimal hyperparameters selected for the two Random Forest models are summarized in Table~S2, and the corresponding cross-validation metrics after tuning are reported in Table~\ref{tab:tuned_test_metrics}. Hyperparameter optimization yields only modest improvements in performance for both $\kappa_\mathrm{L}$ and $\kappa_\mathrm{e}$ predictions. Nevertheless, the tuned models show test-set performance (Figure~\ref{fig:test_performances} and Table~\ref{tab:tuned_test_metrics}) that is consistent with the cross-validation results, indicating that hyperparameter optimization did not lead to overfitting. The parity plots in Figure~\ref{fig:test_performances} further show that both models perform particularly well for materials with low thermal conductivity, which constitute the majority of the training data from the Starrydata repository and are characteristic of thermoelectric materials.

\begin{table}[htbp]
    \centering
    \small
    \caption{\textbf{Performance metrics of the fine-tuned Random Forest models.} Comparison of the predictive accuracy during cross-validation (post-hyperparameter tuning) and on the independent held-out test set for the lattice ($\kappa_{\mathrm{L}}$) and electronic ($\kappa_{\mathrm{e}}$) thermal conductivity models. Units for MAE and RMSE are W\,m$^{-1}$\,K$^{-1}$.}
    \label{tab:tuned_test_metrics}.
    \label{tab:tuned_test_metrics}
    \begin{tabular}{lC{1.4cm}C{1.4cm}C{1.4cm}C{1.4cm}C{1.4cm}C{1.4cm}}
        \toprule
        \multirow{2}{*}{Model} & \multicolumn{3}{c}{Cross-validation} & \multicolumn{3}{c}{Held-out test} \\
        \cmidrule(lr){2-4}\cmidrule(lr){5-7}
        & MAE & RMSE & R$^2$ & MAE & RMSE & R$^2$ \\
        \midrule
        Lattice thermal conductivity ($\kappa_{\mathrm{L}}$)      & 0.35 & 0.62 & 0.80 & 0.34 & 0.61 & 0.80 \\
        Electronic thermal conductivity ($\kappa_{\mathrm{e}}$)   & 0.22 & 0.38 & 0.75 & 0.20 & 0.36 & 0.75 \\
        \bottomrule
    \end{tabular}
\end{table}

While separately predicting $\kappa_{\mathrm{L}}$ and $\kappa_{\mathrm{e}}$ provides more granular insight into a material’s thermal conductivity, it raises the question of whether this decomposition degrades the accuracy of the resulting $\kappa$. This is particularly relevant because our approach computes $\kappa$ as $\kappa_{\mathrm{L}} + \kappa_{\mathrm{e}}$, so errors from the two models could in principle accumulate, leading to worse MAE and RMSE than a single model trained directly on total thermal conductivity. To assess this, we compared $\kappa$ obtained by summing the predictions of the $\kappa_{\mathrm{L}}$ and $\kappa_{\mathrm{e}}$ models with $\kappa$ predicted by a baseline model trained directly on total thermal conductivity (details in the Supplementary Information). The parity plots in Figure~S6 show that the additive approach achieves an R$^2$ above 0.8 on the held-out test set, fully matching the performance fidelity of the direct baseline model. Given this robust performance, we use $\kappa = \kappa_{\mathrm{L}} + \kappa_{\mathrm{e}}$ computed from the separate models for all subsequent analyses.

The observed variance in our optimized models fundamentally reflects the theoretical limit of relying solely on elemental-level chemical descriptors (the Magpie feature set) and measurement temperature. In reality, thermal conductivity is also highly sensitive to crystal structure, defects, and microstructural characteristics that are not captured in the present feature space. This is further illustrated in Figure~S7, which examines the model's predictions for the well-studied compound Sb$_2$Te$_3$. Even at the same temperature, the literature reports a wide spread of thermal conductivity values for Sb$_2$Te$_3$ prepared via different synthesis routes~\cite{saleemi2014evaluation,hu2015enhanced,yang2015facile,dong2010microwave}. While our model's predictions successfully capture both the average magnitude and the temperature dependence of the lattice and carrier thermal conductivity across these Sb$_2$Te$_3$ samples, the model cannot resolve variations beyond composition and temperature that arise from structural and microstructural differences, a behavior also observed in other composition-based models~\cite{barua2025machine}.

The primary reason for excluding these additional structure-related descriptors is not merely the lack of recorded information in the Starrydata repository, but also the inherent difficulty of extracting reliable microstructural data from the literature. Thermoelectric materials typically contain complex, multi-element compositions with varying dopants or alloying elements, making automated, rule-based extraction of key compositional and structural attributes from raw chemical formulas alone extremely challenging. {Although state-of-the-art large language models, such as GPT, offer a promising route for extracting doping or structural information from the literature~\cite{dagdelen2024structured,itani2025large}, these tools are ultimately handcuffed by the sparsity and ambiguity of the underlying source data. In standard experimental practice, comprehensively refined Crystallographic Information Files (CIFs) are rarely published for every specific doped derivative. For example, Itani et al.~\cite{itani2025large} found that more than half of thermoelectric-related papers in their 100-paper sample lacked basic structural features such as space groups or lattice parameters. While undoped prototypes exist in databases like the OQMD and Materials Project, using them as proxies for highly defective experimental samples introduces fatal physical artifacts. For these reasons, we rely on composition-only descriptors and temperature in the present work. This feature space proves highly effective for the first-pass macroscopic screening of $\kappa_{\mathrm{L}}$, $\kappa_{\mathrm{e}}$, and $\kappa$, serving to confidently down-select PGEC candidates for rigorous downstream validation via experiment or first-principles simulations.

\subsection{Feature Importance Analysis}
Beyond the practical necessity of deriving the PGEC descriptor ($\kappa_{\mathrm{L}}/\kappa$), the fundamental physical motivation for modeling $\kappa_{\mathrm{L}}$ and $\kappa_{\mathrm{e}}$ separately is that phonon and electron heat transport are governed by entirely distinct microscopic mechanisms. This distinction is evident in Figure~\ref{fig:shap_results}, where we apply SHAP analysis to probe what the models have actually learned and use it to gain chemical insights into the factors that control each transport channel.

\begin{figure}[htbp]
    \centering
    \includegraphics[width=0.6\linewidth]{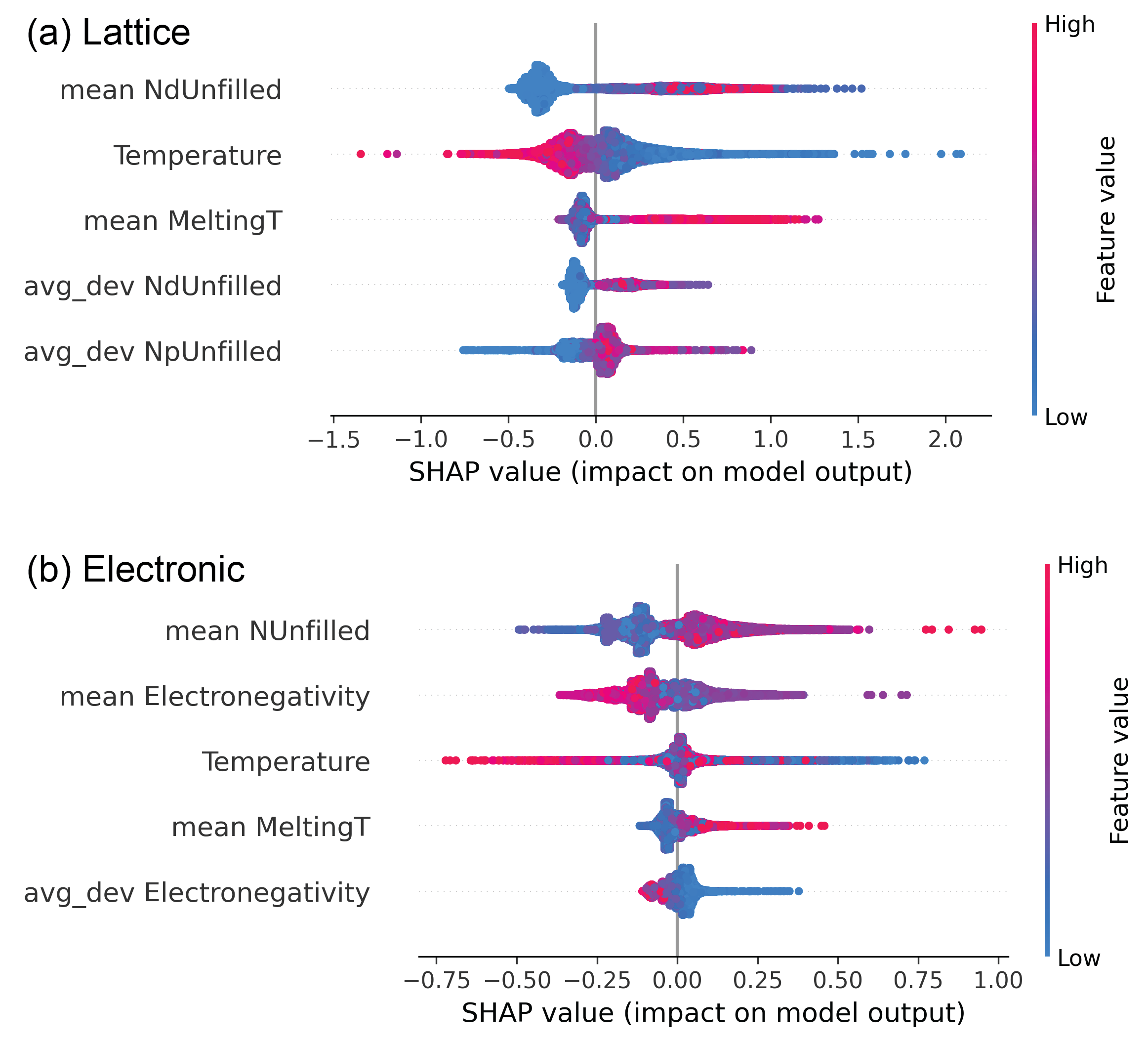}
    \caption{\textbf{Feature importance and chemical interpretability of the dual machine learning models.} SHapley Additive exPlanations (SHAP) beeswarm plots for the \textbf{a} $\kappa_{\mathrm{L}}$ and \textbf{b} $\kappa_{\mathrm{e}}$ Random Forest models. Features are ranked top-to-bottom by their mean absolute SHAP value, illustrating how specific global and local compositional descriptors distinctively impact the predicted thermal transport components.}
    \label{fig:shap_results}
\end{figure}

In both models, \texttt{Temperature} naturally emerges as a dominant feature, with higher $T$ driving lower predictions for both transport channels. This aligns perfectly with physical expectations: increasing temperature strongly enhances both phonon–phonon (Umklapp) scattering and electron–phonon scattering. These intensified scattering events fundamentally suppress lattice heat transport ($\kappa_{\mathrm{L}}$) and carrier mobility, thereby reducing $\sigma$ and correspondingly suppressing $\kappa_{\mathrm{e}}$ via the Wiedemann–Franz relation. However, for $\kappa_{\mathrm{e}}$, the SHAP plot also reveals a subset where the model's response to elevated $T$ is either muted or slightly positive. We attribute this dispersion to non-degenerate semiconductor samples or inherent mathematical competition within the Wiedemann–Franz relation ($\kappa_{\mathrm{e}} = L \sigma T$): while $\sigma$ typically decays with temperature, the explicit temperature multiplier ($T$) and the fluctuations in the Lorenz number ($L$, through its correlation with $S$) can occasionally offset this reduction. Another consistently important input is \texttt{mean MeltingT}, which tends to increase both $\kappa_{\mathrm{L}}$ and $\kappa_{\mathrm{e}}$. High melting points generally correlate with strong bonding, which is associated with larger phonon group velocities and higher electron densities. As a result, materials with larger \texttt{mean MeltingT} tend to exhibit higher predicted $\kappa_{\mathrm{L}}$ and $\kappa_{\mathrm{e}}$.

For the $\kappa_{\mathrm{L}}$ model, the unfilled $d$-electron count, \texttt{mean$\,$NdUnfilled}, has an overall positive effect on $\kappa_{\mathrm{L}}$, meaning that compositions enriched in $s$-block and early $d$-block elements are predicted to have higher $\kappa_{\mathrm{L}}$. Chemically, these elements favor the formation of strong and directional $s$–$p$ and $d$–$p$ hybridized bonds that stiffen the lattice and increase phonon group velocities. Conversely, late $d$-block elements, such as Cu and Ag, with fuller $d$ shells, tend to form softer and more anharmonic bonds that hinder phonon transport. Notably, a few compositions with low \texttt{mean$\,$NdUnfilled} still show high $\kappa_{\mathrm{L}}$. these likely correspond to $p$-block–dominated, super-hard networks, such as AlN and SiC, which also possess strong bonding and thus high lattice thermal conductivity. 
Furthermore, the average deviation of unfilled $d$- and $p$-electron counts (\texttt{avg$\,$dev$\,$NdUnfilled} and \texttt{avg$\,$dev$\,$NpUnfilled}) positively influence $\kappa_{\mathrm{L}}$. These features quantify the contrast in unfilled $d$/$p$ counts between constituent elements, that is, the degree of ``role differentiation'' or bonding complementarity between cations and anions. Larger contrast typically accompanies strong ionic--covalent hybridization and high bond stiffness, which promote higher phonon group velocity and thus higher $\kappa_{\mathrm{L}}$.

In contrast, the $\kappa_{\mathrm{e}}$ model highlights features that act as proxies for electron affinity and macroscopic metallicity. First, the mean unfilled valence-orbital count, \texttt{mean$\,$NUnfilled}, has a positive impact on $\kappa_{\mathrm{e}}$, since more unfilled valence states favor electronic delocalization, increase metallicity and $\sigma$, and therefore boost $\kappa_{\mathrm{e}}$. Second, \texttt{mean$\,$Electronegativity} shows a negative impact on $\kappa_{\mathrm{e}}$. Lower mean electronegativity corresponds to more metallic bonding that donates conduction electrons, enhancing $\sigma$ and $\kappa_{\mathrm{e}}$, whereas higher mean electronegativity reflects stronger electron affinity and more localized electrons, which suppress carrier mobility and reduce $\kappa_{\mathrm{e}}$. Finally, \texttt{avg$\,$dev$\,$Electronegativity} also impacts $\kappa_{\mathrm{e}}$ negatively. Larger inter-element electronegativity mismatch implies stronger charge transfer and more polar bonding, promoting electron localization and lowering carrier mobility, which in turn decreases $\kappa_{\mathrm{e}}$. Small electronegativity contrast (near-isoelectronegativity) instead favors delocalization and more conductive network, increasing $\kappa_{\mathrm{e}}$. 

Overall, these statistically derived trends are in line with our chemical intuition, indicating that both models capture reasonable relationships between composition and lattice/carrier thermal conductivity. By jointly interpreting the $\kappa_{\mathrm{L}}$ and $\kappa_{\mathrm{e}}$ models, we obtain a chemically transparent picture of how phonon and electron transport can be differentiated and, to some extent, decoupled in thermoelectric materials. In particular, “global’’ levers, such as overall bond strength (\texttt{mean$\,$MeltingT}), intrinsically couple the two properties, driving both $\kappa_{\mathrm{L}}$ and $\kappa_{\mathrm{e}}$ in the same direction. Therefore, optimizing these features alone cannot realize the PGEC ideal.
Breaking this correlation requires targeting levers that act selectively on a single channel.To explicitly suppress $\kappa_{\mathrm{L}}$, the lattice must be chemically softened by incorporating weak-bonding, highly anharmonic atoms with minimal cation–anion valence disparities. Meanwhile, increasing $\kappa_{\mathrm{e}}$ requires metallic, low-electronegativity-contrast environments that support electron exchange, band formation, and long-range delocalization even in a relatively soft lattice. Mapped onto the periodic table, these requirements are most naturally satisfied by combinations of late $d$-block and heavy $p$-block elements that decouple soft bonding from high electron density. Therefore, the core chemical insight from our data-driven interpretation is that promising PGEC thermoelectrics are those in which elemental combinations decouple bond strength from electron density, enabling a phonon-glass lattice to coexist with an electron-crystal conduction network.

\section{Discussion}
\subsection{Material Screening}
Having established that the $\kappa_{\mathrm{L}}$ and $\kappa_{\mathrm{e}}$ models are reasonably accurate and chemically interpretable, we proceed to use them for the data-driven screening and PGEC optimization of uncharacterized thermoelectric materials. Specifically, we deploy the trained models to 104{,}567 unique compositions from the Materials Project database~\cite{jain2013commentary} that are not present in the training or test sets. As shown in Figure~\ref{fig:findings_1}(a), our initial screening focuses on binary and ternary compounds that are predicted to be thermodynamically stable (energy above hull smaller than 0.1~eV/atom) and exhibit finite band gaps. Among the 14{,}830 compounds that satisfy these criteria, our dual-model framework identifies 2{,}522 materials with low $\kappa$ ($\leq 2$~\wmk) at 300~K, marking them as promising candidates for further exploration. The predicted $\kappa$ and $\kappa_\mathrm{L}/\kappa$ values for these 2{,}522 compositions are shown in Figure~\ref{fig:findings_1}(b).

\begin{figure}[htbp]
    \centering
    \includegraphics[width=\linewidth]{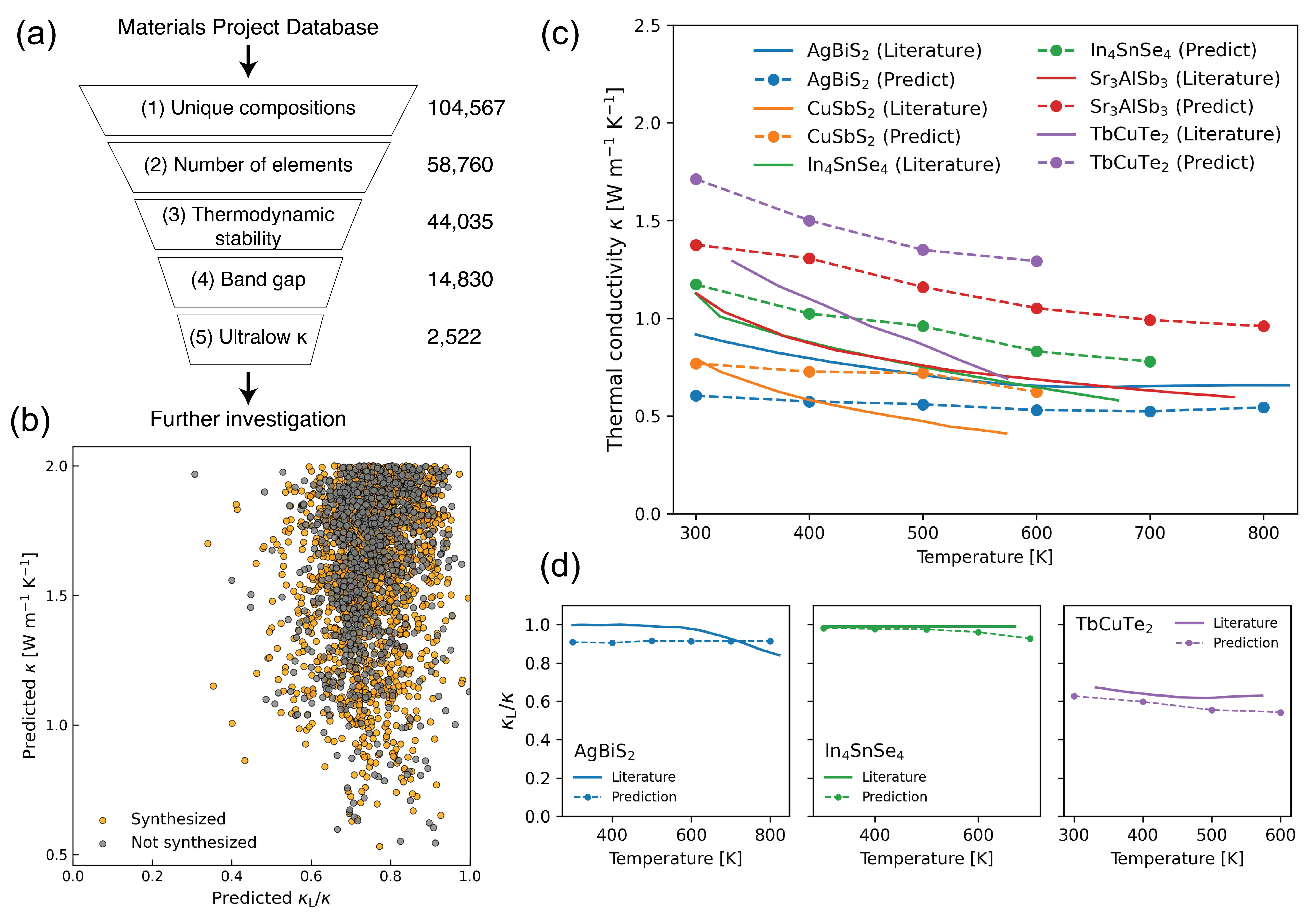}
    \caption{\textbf{High-throughput materials screening and out-of-distribution experimental validation.} \textbf{a} The sequential data-driven workflow applied to the Materials Project database to screen for thermodynamically stable, pristine compositions with ultralow total thermal conductivity ($\kappa \le 2$~W\,m$^{-1}$\,K$^{-1}$) at 300~K. \textbf{b} Distribution of total $\kappa$ versus the PGEC descriptor ($\kappa_{\mathrm{L}}/\kappa$) for the 2{,}522 positively identified candidates. \textbf{c}, \textbf{d} Comparison between our model predictions (dashed lines) and historically reported literature values (solid lines) for temperature-dependent total $\kappa$ (\textbf{c}) and $\kappa_{\mathrm{L}}/\kappa$ (\textbf{d}) across five unseen experimental compounds.}
    \label{fig:findings_1}
\end{figure}

To further evaluate our models' generalizability to unseen real-world TE materials, we randomly selected, from this set of 2{,}522 compounds, several reported but historically overlooked materials, including AgBiS$_2$~\cite{AgBiS2}, CuSbS$_2$~\cite{CuSbS2}, In$_4$SnSe$_4$~\cite{In4SnSe4}, Sr$_3$AlSb$_3$~\cite{Sr3AlSb3}, and TbCuTe$_2$~\cite{TbCuTe2}. We verified that neither these pristine host compounds nor any of their doped variants exist in our curated dataset. For each blind candidate, we compare the predicted (dashed lines) and literature-reported (solid lines) temperature-dependent $\kappa$ in Figure~\ref{fig:findings_1}(c). The overall agreement is reasonably good, and the temperature-dependent trends of all five materials are also well captured. 

Additionally, explicitly reported $\kappa_{\mathrm{L}}$ values are available in the literature for AgBiS$_2$, In$_4$SnSe$_4$, and TbCuTe$_2$. This presents an opportunity to directly validate our PGEC descriptor, $\kappa_{\mathrm{L}}/\kappa$. As demonstrated in Figure~\ref{fig:findings_1}(d), the predicted $\kappa_\mathrm{L}/\kappa$ values are also in good agreement with the reported values. Importantly, our framework correctly categorizes whether each pristine compound lies above or below the critical $\kappa_\mathrm{L}/\kappa$ = 0.5 threshold, which we propose to separate different optimization strategies for driving materials toward the PGEC regime and thereby enhancing $ZT$. Together, these results demonstrate how the two metrics, $\kappa$ and $\kappa_\mathrm{L}/\kappa$, can be leveraged by machine learning models to screen previously unseen compositions for not only ultralow but also PGEC-favorable thermal conductivity.

\subsection{Material Optimization}

In practice, thermoelectric materials design requires not only identifying promising candidates but also predicting how their performance shifts under chemical doping. To explore whether our models can qualitatively guide such optimization toward the PGEC regime, we selected AgBiS$_2$ as a case study. For pristine AgBiS$_2$, it already displays an ultralow total thermal conductivity close to 0.5\wmk, but $\kappa_\mathrm{L}/\kappa$ remains above 0.8, far from the PGEC optimum, indicating that injecting carriers to boost electronic thermal transport is the priority. Given that AgBiS$_2$ is an intrinsic $n$-type semiconductor, we deployed our models to predict $\kappa$ and $\kappa_\mathrm{L}/\kappa$ for a series of rational $n$-type dopants at both the Ag and S sites, as illustrated in Figure~\ref{fig:findings_2}. All compositional modifications are fixed at 10~at.\% per site and predictions are performed at 800~K, the precise temperature regime where pristine AgBiS$_2$ attains its peak $ZT$ in the literature~\cite{AgBiS2}.

\begin{figure}[htbp]
    \centering
    \includegraphics[width=0.5\linewidth]{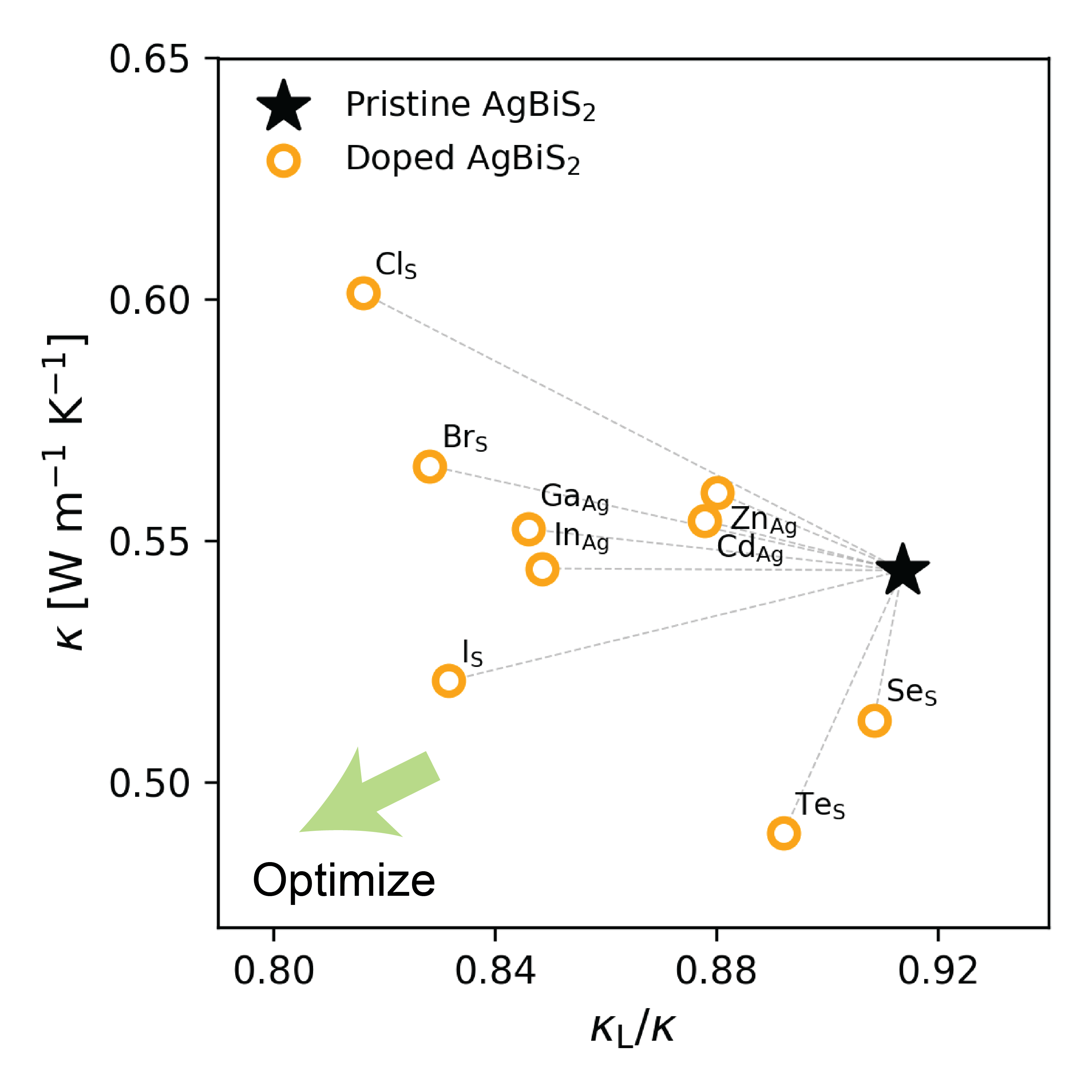}
    \caption{\textbf{Data-driven PGEC optimization trajectories for $n$-type AgBiS$_2$.} The plot maps the predicted shifts in total thermal conductivity ($\kappa$) and the lattice-to-total ratio ($\kappa_{\mathrm{L}}/\kappa$) upon introducing various $n$-type dopants (10~at.\%) at the Ag and S sites at 800~K. The black star denotes the pristine AgBiS$_2$ host, while open circles represent the chemically modified variants. The green arrow visually underscores the fundamental PGEC design objective: simultaneously suppressing total $\kappa$ while actively driving the highly unbalanced $\kappa_{\mathrm{L}}/\kappa$ ratio down toward the optimal $\approx$0.5 regime.}
    \label{fig:findings_2}
\end{figure}

Among the S-site modifications, aliovalent halogen substitutions (Cl, Br, I) all successfully enhance $\kappa_\mathrm{e}$ and thereby drive $\kappa_{\mathrm{L}}/\kappa$ downward, consistent with their role as electron donors. In particular, the predicted effect of Cl$_\mathrm{S}$ is in good qualitative agreement with the experimental 
findings of An et al.~\cite{AgBiS2}, who showed that Cl substitution 
effectively enhances the carrier concentration and decreases \klk from 
approximately 0.84 to 0.80 at 823~K. However, the enhanced 
$\kappa_\mathrm{e}$ also increased total $\kappa$, leading to only a minor $ZT$ improvement (to 0.45). Importantly, our predictions suggest that heavier halogens, Br and I, may produce a smaller $\kappa_\mathrm{e}$ enhancement but simultaneously suppress $\kappa_\mathrm{L}$, resulting in a lower 
total $\kappa$. This synergistic effect of enhancing electronic transport while reducing lattice thermal conductivity makes Br and I highly strategic candidates for subsequent experimental realization. In contrast, isovalent Se and Te substitutions on the S site, primarily reduce $\kappa_\mathrm{L}$ with negligible impact on $\kappa_\mathrm{e}$, lowering total $\kappa$ but leaving $\kappa_{\mathrm{L}}/\kappa$ largely unchanged, consistent with their role as phonon-scattering centers rather than carrier donors. For the Ag-site substitutions, elements like Zn and Cd (each donate one extra electron) are predicted to induce a moderate enhancement in $\kappa_\mathrm{e}$, while In, Ga (donating two extra electrons) produce correspondingly larger enhancement. However, compared to S-site halogen doping, the Ag-site dopants are predicted to be less effective at lowering $\kappa_{\mathrm{L}}/\kappa$. Experimentally, An et al.~\cite{AgBiS2} attributed this to the intrinsic Ag/Bi cation disorder in AgBiS$_2$, which suppresses the effectiveness of cation-site doping on electrical transport. 

Overall, these strategic predictions demonstrate that our composition-based framework can effectively leverage the proposed pair of metrics, $\kappa$ and $\kappa_{\mathrm{L}}/\kappa$ to move beyond static screening, providing actionable, PGEC-oriented doping pathways. However, we must candidly acknowledge the inherent limitations of deploying a Magpie-based model for $\kappa_{\mathrm{e}}$ predictions. For $\kappa_{\mathrm{L}}$, composition-averaged descriptors are well-suited, as lattice thermal conductivity is largely governed by properties that Magpie captures effectively, such as mass contrast, atomic size mismatch, and bonding stiffness. For $\kappa_{\mathrm{e}}$, however, the carrier concentration depends critically on site-specific information. Whether a dopant acts as a donor or acceptor strictly depends on the specific crystallographic host site it occupies. This topological resolution is intrinsically absent from composition-averaged descriptors. More fundamentally, this barrier is perpetuated by existing databases, which rarely standardize or record explicit substitution sites and dopants, stalling the field's transition beyond bulk compositional featurization. As a result, while our models capture qualitative doping trends in many cases, they fundamentally treat the doped matrix as a homogenized solid solution, rendering the $\kappa_{\mathrm{e}}$ predictions somewhat blunted to precise local defect mechanics. Nevertheless, the underlying physical compass, the $\kappa_{\mathrm{L}}/\kappa$ PGEC descriptor, remains profoundly robust. We anticipate that as future, site-aware foundation models emerge, they will readily adopt this exact PGEC descriptor to provide high-fidelity optimization guidance, definitively closing the critical loop between computational materials screening and targeted performance maximization.

\subsection{Summary and outlook}
In this work, we used a massive, experimentally derived thermoelectric dataset to reconceptualize the phonon-glass electron-crystal (PGEC) concept from a qualitative ideal into a quantitative, data-driven design rule. By systematically evaluating the lattice-to-total thermal conductivity ratio \( \kappa_\mathrm{L}/\kappa \) across tens of thousands of measurements, we demonstrated that high-\( ZT \) materials not only reside in the low $\kappa$ regime, but also cluster around \( \kappa_\mathrm{L}/\kappa \approx 0.5 \). This insightful observation establishes that an approximately equal partition of thermal transport between phonons and charge carriers serves as a semi-quantitative signature of the PGEC regime---a universal trend rigorously validated across two representative thermoelectric families, CoSb$_3$ and GeTe.

Guided by this insight, we deployed two composition- and temperature-based machine learning models to decouple $\kappa_{\mathrm{L}}$ and $\kappa_{\mathrm{e}}$, enabling the simultaneous evaluation of both $\kappa$ and \( \kappa_\mathrm{L}/\kappa \) in high-throughput screening. Subsequent SHAP analysis of both models provided chemically intuitive guidelines for element selection that balance electrical and phonon transport, linking specific compositional motifs to soft lattices and delocalized electron density. We then successfully validated the framework's general predictive fidelity on a set of recently reported, underexplored compounds. Finally, using AgBiS$_2$ as a case study, we highlight how the two metrics, $\kappa$ and the PGEC descriptor \( \kappa_\mathrm{L}/\kappa \), can be leveraged in a data-driven framework to go beyond static material screening and provide reasonable optimization routes to tune \( \kappa_\mathrm{L}/\kappa \) toward the PGEC target of 0.5 for improved thermoelectric performance. Our approach bridges the critical gap between computational screening and experimental realization, ultimately laying a robust, AI-accelerated foundation for the next generation of rational thermoelectric materials design.

\section{Methods}
\subsection{Data curation}
We used the 2025-07-01 version of the Starrydata database~\cite{katsura2025starrydata}, which contains 202{,}178 digitized curves (an excerpt is shown in Table~\ref{tab:starrydata_excerpt}). From these records, we retained samples (identified by sample-id) that reported a complete set of thermoelectric properties: $S$, $\sigma$, $\kappa$, and $ZT$. We then removed samples with invalid compositions or duplicate curves for any of these properties as a basic quality-control step. After confirming unit consistency across the retained records, we removed entries with physically implausible values, such as negative $T$, $\kappa$, or $\sigma$. We also removed entries with $ZT > 3$, since such values are extraordinarily rare for bulk thermoelectrics and too sparse to meaningfully inform model training, and often reflect human input errors during the database compilation.

\begin{table}[htbp]
  \centering
  \small
  \caption{\textbf{Excerpt of the raw Starrydata records.} }
  \label{tab:starrydata_excerpt}
  \begin{tabular}{llllllll}
    \toprule
    Sample-id & Composition & Prop-\texttt{x} & Prop-\texttt{y} & Unit-\texttt{x} & Unit-\texttt{y} & \texttt{x} & \texttt{y} \\ \midrule
    5422 & In$_{0.3}$Co$_4$Sb$_{12}$  & $T$ & $S$ & K & $\mu$V\,K\textsuperscript{-1} & [323.3, 418.7, 518.5] & [-163.3, -185.0, -209.5] \\
    5422 & In$_{0.3}$Co$_4$Sb$_{12}$ & $T$ & $\kappa$ & K & W\,m\textsuperscript{$-$1}\,K\textsuperscript{$-$1} & [324.3, 423.1, 521.9] & [3.50, 3.11, 2.83] \\
    5422 & In$_{0.3}$Co$_4$Sb$_{12}$ & $T$ & $\sigma$ & K & S\,cm\textsuperscript{$-$1} & [318.5, 422.4, 519.4] & [1416.0, 1192.0, 1034.0]\\
    5422 & In$_{0.3}$Co$_4$Sb$_{12}$ & $T$ & $ZT$ & K & - & [320.7, 421.7, 524.6] & [0.35, 0.57, 0.84] \\ \bottomrule
  \end{tabular}
\end{table}

In manually curated databases like Starrydata, thermoelectric properties of the same sample-id are not perfectly aligned at the same temperature, since they are extracted from published plots by humans using digitization software. Small mismatches can occur during the click-and-extract process, as exemplified in the \texttt{x} column of Table~\ref{tab:starrydata_excerpt}. These slight temperature offsets across properties make it impossible to directly calculate $\kappa_\mathrm{e}$ and $\kappa_\mathrm{L}$ from $S$, $\sigma$, and $\kappa$. To resolve this, we first performed temperature alignment using the reported $ZT$ temperature as the anchor: for each reported $ZT$ at a given temperature $T$, we searched for $S$, $\sigma$, and $\kappa$ values for the same sample-id within a $\pm 5$~K window around $T$. If multiple values were found, the one closest to $T$ was chosen. This window is small enough that variations in the properties are typically negligible.

To further ensure internal consistency and eliminate possible labeling errors, we recomputed $ZT$ at each reported $ZT$ temperature using the corresponding $S$, $\sigma$, and $\kappa$ values, which we denote $ZT_{\mathrm{calc}}$. By definition, if all thermoelectric properties are correctly extracted from the published figures, $ZT_{\mathrm{calc}}$ should match the author-reported $ZT$. Therefore, we excluded entries for which $ZT_{\mathrm{calc}}$ deviated from $ZT$ by more than 10\%. This threshold was chosen as it strictly filters out significant digitization artifacts while generously accommodating typical rounding differences and standard experimental measurement uncertainties. This rigorous ``$ZT$ consistency'' check is also the main reason for retaining only samples with a complete set of thermoelectric properties, as mentioned earlier.

Table~\ref{tab:starrydata_explode} shows the temperature-aligned data corresponding to Table~\ref{tab:starrydata_excerpt}. In this example, temperature alignment fails at $T = 524.6$~K because the closest Seebeck coefficient data (at $T = 518.5$~K) lies outside the $\pm 5$~K window, so that row is flagged for removal. After temperature alignment, we computed $ZT_{\mathrm{calc}}$ and confirmed that the remaining two entries at $T = 320.7$ and 421.7~K show consistent $ZT_{\mathrm{calc}}$ and $ZT$. Some examples of removed records with a relative difference from the reported $ZT$ exceeding 10\% are provided in the Supplementary Information (Table~S1). After this initial round of data cleaning, 89,859 records remain, each representing a set of TE properties measured at a specific composition and temperature.

\begin{table}[htbp]
  \centering
  \small
  \caption{\textbf{Temperature-aligned thermoelectric properties and consistency validation.} This table demonstrates the result of applying the $\pm 5$~K temperature alignment window to the raw records from Table~\ref{tab:starrydata_excerpt}.}
  \label{tab:starrydata_explode}
  \begin{tabular}{lccccccc}
    \toprule
    \multirow{2}{*}{Sample-id} & \multirow{2}{*}{Composition} & $T$ & $S$ & $\kappa$ & $\sigma$ & \multirow{2}{*}{$ZT$} & \multirow{2}{*}{$ZT_{\mathrm{calc}}$} \\
    & & (K)& ($\mu$V\,K$^{-1}$) & (\wmk) & (S\,cm$^{-1}$) & & \\
    \midrule
    5422 & In$_{0.3}$Co$_4$Sb$_{12}$ & 320.7 & -163.3 & 3.50 & 1416.0 & 0.35 & 0.35  \\
    5422 & In$_{0.3}$Co$_4$Sb$_{12}$ & 421.7 & -185.0 & 3.11 & 1192.0 & 0.57 & 0.55  \\
    5422 & In$_{0.3}$Co$_4$Sb$_{12}$ & 524.6 & - & 2.83 & 1034.0 & 0.84 & -  \\ \bottomrule
  \end{tabular}
\end{table}

\subsection{Lattice and carrier thermal conductivity calculations}
To avoid data skewness and focus on the practical operating window, we further restricted the dataset to the range $300 \le T \le 1000$~K and $\kappa \le 10$~W\,m$^{-1}$\,K$^{-1}$. This window corresponds to the temperature and thermal conductivity range most commonly studied in thermoelectric research and incorporates nearly all of the curated data. Total thermal conductivity $\kappa$ was then split into lattice thermal conductivity ($\kappa_{\mathrm{L}}$) and carrier thermal conductivity ($\kappa_{\mathrm{e}}$) using the Wiedemann–Franz law:
\begin{align}
    \kappa_{\mathrm{L}} &= \kappa - \kappa_{\mathrm{e}},\\
    \kappa_{\mathrm{e}} &= L \,\sigma\, T,
\end{align}
where $L$ is the Lorenz number (in $10^{-8}\ \mathrm{W\,\Omega\,K^{-2}}$), estimated from $S$ (in $\mu$V\,K$^{-1}$) using:
\begin{equation}
    L=1.5+\mathrm{exp}[-\frac{|S|}{116}]
\end{equation}

This approximation is typically accurate within $\sim$5\% for semiconductors satisfying the single parabolic band (SPB) model and within $\sim$20\% for more complex systems~\cite{kim2015characterization}. To further assess the accuracy of our calculated $\kappa_{\mathrm{e}}$, we compared them against a subset of 27{,}140 Starrydata entries containing literature-reported $\kappa_{\mathrm{e}}$ values in Figure~S1. Overall, we observed excellent agreement between our calculated $\kappa_{\mathrm{e}}$ and the literature values, with $R^2 = 0.96$ and a mean absolute difference of 0.09 W\,m$^{-1}$\,K$^{-1}$. 

Notably, we found that a substantial fraction of the discrepancy originates from oversimplifications in earlier works that use approximate, constant Lorenz numbers~\cite{liu2012copper,xue2020off,wang2011enhanced,sun2016effect,harnwunggmoung2010high,kim2019enhanced,tanusilp2020beneficial,xu2021substantial}. The conclusions of some studies on thermal conductivity could change significantly if their analyses were updated using our $S$-dependent Lorenz numbers, which further emphasizes the importance of constructing a uniform and consistent dataset. Nevertheless, even in cases where our calculated thermal conductivities match well with the reported values, there are anomalous cases where $\kappa_{\mathrm{L}}$ or $\kappa_{\mathrm{e}}$ approaches zero. These anomalies can be attributed to breakdowns of the SPB model due to phase impurities~\cite{el2021unraveling,liang2021ductile}, phase metastability and transitions~\cite{dong2019medium}, or extremely complex transport behavior~\cite{falkenbach2015tin}. Given the rarity of these cases and the absence of a universally applicable alternative method to parse $\kappa_{\mathrm{L}}$ and $\kappa_{\mathrm{e}}$ across all materials, we conservatively retained these data points for future investigation. A more detailed discussion of the uncertainties propagated into $\kappa_{\mathrm{L}}$ and $\kappa_{\mathrm{e}}$ is provided in the Supplementary Information.

\subsection{Machine learning framework}
The final curated dataset contains 71{,}913 entries, which we split into training (80\%) and test (20\%) sets while avoiding composition overlap. Specifically, we perform a group-based split so that all entries sharing the same reduced formula are assigned to the same set, preventing any reduced formula from appearing in both sets. Figure~S2 shows examples of the temperature-dependent $ZT$ and $\kappa$ in the final dataset.

Each unique composition in the curated dataset was featurized using its Magpie chemical descriptors~\cite{ward2016general} alongside the measurement temperature, yielding an initial set of 133 features for predicting $\kappa_{\mathrm{L}}$ and $\kappa_{\mathrm{e}}$. Features with zero variance across the training data were first removed, followed by a Pearson correlation analysis on the remaining features. For any pair with an absolute correlation coefficient greater than 0.99, one of the two features was discarded. The correlated feature pairs and the features removed are detailed in the Supplementary Information. Feature removal was performed exclusively on the training set to prevent information leakage, and the resulting feature set was then applied to the test set. The final feature set used for training contains 108 features in total, consisting of 107 Magpie descriptors and the temperature.

We evaluated four tree-based algorithms (XGBoost, LightGBM, Random Forest, and CatBoost), chosen for their well-documented capability to capture highly non-linear composition-property relationships in thermoelectric materials~\cite{parse2023machine,wang2020identification,barua2024interpretable}. Model performance was assessed using 5-fold GroupKFold cross-validation, grouping by reduced formula to ensure no composition overlap among folds. The primary metric for model selection was the mean absolute error (MAE), chosen for its interpretability and reduced sensitivity to outliers. The model with the lowest cross-validated MAE was then further tuned using GridSearchCV, retrained on the full training set, and finally evaluated on the independent test set. Furthermore, to provide a rigorous baseline for performance comparison, we trained an equivalent single model tasked with directly predicting $\kappa$. As detailed in the Supplementary Information, comparing our dual-model predictions against this baseline confirms that our physically decoupled approach (predicting $\kappa_{\mathrm{L}}$ and $\kappa_{\mathrm{e}}$ independently) successfully provides granular insights into the underlying transport mechanisms without compromising the overall prediction accuracy for $\kappa$.

\section*{Funding}
This work was supported by MEXT Innovative Nuclear
Research and Development Program Grant Number JPMXD0220354330 and JPMXD0222682541

\section*{Author contributions}
Y.S. and Z.L. conceived the research ideas. Y.S. and T.I. curated the dataset. Z.L. led the thermoelectric trend and data overview analysis. Y.S. performed additional data analysis and trained the machine learning models. Y.S. and Z.L. carried out the materials screening and optimization, and drafted the manuscript. Y.O., C.W., and K.K. supervised the research. K.K. acquired the funding. All authors contributed to discussions and provided feedback on the manuscript.

\bibliographystyle{unsrtnat}
\bibliography{references}  

@article{forman2016estimating,
  title={Estimating the global waste heat potential},
  author={Forman, Clemens and Muritala, Ibrahim Kolawole and Pardemann, Robert and Meyer, Bernd},
  journal={Renewable and Sustainable Energy Reviews},
  volume={57},
  pages={1568--1579},
  year={2016},
  publisher={Elsevier}
}

@article{geffroy2021techno,
  title={Techno-economic analysis of waste-heat conversion},
  author={Geffroy, Charles and Lilley, Drew and Parez, Pedro Sanchez and Prasher, Ravi},
  journal={Joule},
  volume={5},
  number={12},
  pages={3080--3096},
  year={2021},
  publisher={Elsevier}
}

@article{bell2008cooling,
  title={Cooling, heating, generating power, and recovering waste heat with thermoelectric systems},
  author={Bell, Lon E},
  journal={Science},
  volume={321},
  number={5895},
  pages={1457--1461},
  year={2008},
  publisher={American Association for the Advancement of Science}
}

@article{snyder2008complex,
  title={Complex thermoelectric materials},
  author={Snyder, G Jeffrey and Toberer, Eric S},
  journal={Nature Materials},
  volume={7},
  number={2},
  pages={105--114},
  year={2008},
  publisher={Nature Publishing Group UK London}
}

@article{ward2016general,
  title={A general-purpose machine learning framework for predicting properties of inorganic materials},
  author={Ward, Logan and Agrawal, Ankit and Choudhary, Alok and Wolverton, Christopher},
  journal={npj Computational Materials},
  volume={2},
  number={1},
  pages={1--7},
  year={2016},
  publisher={Nature Publishing Group}
}

@article{kim2015characterization,
  title={{Characterization of Lorenz number with Seebeck coefficient measurement}},
  author={Kim, Hyun-Sik and Gibbs, Zachary M and Tang, Yinglu and Wang, Heng and Snyder, G Jeffrey},
  journal={APL Materials},
  volume={3},
  number={4},
  year={2015},
  publisher={AIP Publishing}
}

@article{jain2013commentary,
  title={{Commentary: The Materials Project: A materials genome approach to accelerating materials innovation}},
  author={Jain, Anubhav and Ong, Shyue Ping and Hautier, Geoffroy and Chen, Wei and Richards, William Davidson and Dacek, Stephen and Cholia, Shreyas and Gunter, Dan and Skinner, David and Ceder, Gerbrand and others},
  journal={APL Materials},
  volume={1},
  number={1},
  year={2013},
  publisher={AIP Publishing}
}

@article{parse2023machine,
  title={{Machine learning for predicting ZT values of high-performance thermoelectric materials in mid-temperature range}},
  author={Parse, Nuttawat and Pinitsoontorn, Supree},
  journal={APL Materials},
  volume={11},
  number={8},
  year={2023},
  publisher={AIP Publishing}
}

@article{itani2025large,
  title={{Large language model-driven database for thermoelectric materials}},
  author={Itani, Suman and Zhang, Yibo and Zang, Jiadong},
  journal={Computational Materials Science},
  volume={253},
  pages={113855},
  year={2025},
  publisher={Elsevier}
}

@article{dagdelen2024structured,
  title={Structured information extraction from scientific text with large language models},
  author={Dagdelen, John and Dunn, Alexander and Lee, Sanghoon and Walker, Nicholas and Rosen, Andrew S and Ceder, Gerbrand and Persson, Kristin A and Jain, Anubhav},
  journal={Nature Communications},
  volume={15},
  number={1},
  pages={1418},
  year={2024},
  publisher={Nature Publishing Group UK London}
}

@misc{slack1995crc,
  title={{CRC} handbook of thermoelectrics},
  author={Slack, Glen A and Rowe, DM and others},
  year={1995},
  publisher={CRC press Boca Raton, FL}
}

@article{li2022large,
  title={Large data set-driven machine learning models for accurate prediction of the thermoelectric figure of merit},
  author={Li, Yi and Zhang, Jingzi and Zhang, Ke and Zhao, Mengkun and Hu, Kailong and Lin, Xi},
  journal={ACS Applied Materials \& Interfaces},
  volume={14},
  number={50},
  pages={55517--55527},
  year={2022},
  publisher={ACS Publications}
}

@article{barua2024thermoelectric,
  title={{Thermoelectric material performance (zT) predictions with machine learning}},
  author={Barua, Nikhil K and Lee, Sangjoon and Oliynyk, Anton O and Kleinke, Holger},
  journal={ACS Applied Materials \& Interfaces},
  volume={17},
  number={1},
  pages={1662--1673},
  year={2024},
  publisher={ACS Publications}
}

@article{xu2024prediction,
  title={Prediction of thermoelectric-figure-of-merit based on autoencoder and light gradient boosting machine},
  author={Xu, Yingying and Liu, Xinyi and Wang, Jifen},
  journal={Journal of Applied Physics},
  volume={135},
  number={7},
  year={2024},
  publisher={AIP Publishing}
}

@article{li2025machine,
  title={Machine learning for accelerated prediction of lattice thermal conductivity at arbitrary temperature},
  author={Li, Zihe and Li, Mengke and Luo, Yufeng and Cao, Haibin and Liu, Huijun and Fang, Ying},
  journal={Digital Discovery},
  volume={4},
  number={1},
  pages={204--210},
  year={2025},
  publisher={Royal Society of Chemistry}
}

@article{wang2020identification,
  title={Identification of crystalline materials with ultra-low thermal conductivity based on machine learning study},
  author={Wang, Xinming and Zeng, Shuming and Wang, Zhuchi and Ni, Jun},
  journal={The Journal of Physical Chemistry C},
  volume={124},
  number={16},
  pages={8488--8495},
  year={2020},
  publisher={ACS Publications}
}

@article{barua2024interpretable,
  title={Interpretable machine learning model on thermal conductivity using publicly available datasets and our internal lab dataset},
  author={Barua, Nikhil K and Hall, Evan and Cheng, Yifei and Oliynyk, Anton O and Kleinke, Holger},
  journal={Chemistry of Materials},
  volume={36},
  number={14},
  pages={7089--7100},
  year={2024},
  publisher={ACS Publications}
}

@article{barua2025machine,
  title={Machine Learning Predictions of Thermopower for Thermoelectric Material Screening},
  author={Barua, Nikhil K and Kleinke, Holger},
  journal={ACS Applied Energy Materials},
  year={2025},
  publisher={ACS Publications}
}

@article{yuan2022machine,
  title={Machine learning for accelerated prediction of the Seebeck coefficient at arbitrary carrier concentration},
  author={Yuan, HM and Han, SH and Hu, R and Jiao, WY and Li, MK and Liu, HJ and Fang, Y},
  journal={Materials Today Physics},
  volume={25},
  pages={100706},
  year={2022},
  publisher={Elsevier}
}

@article{na2022public,
  title={A public database of thermoelectric materials and system-identified material representation for data-driven discovery},
  author={Na, Gyoung S and Chang, Hyunju},
  journal={npj Computational Materials},
  volume={8},
  number={1},
  pages={214},
  year={2022},
  publisher={Nature Publishing Group UK London}
}

@article{furmanchuk2018prediction,
  title={{Prediction of seebeck coefficient for compounds without restriction to fixed stoichiometry: A machine learning approach}},
  author={Furmanchuk, Al'ona and Saal, James E and Doak, Jeff W and Olson, Gregory B and Choudhary, Alok and Agrawal, Ankit},
  journal={Journal of Computational Chemistry},
  volume={39},
  number={4},
  pages={191--202},
  year={2018},
  publisher={Wiley Online Library}
}

@article{tewari2020machine,
  title={{Machine learning approaches to identify and design low thermal conductivity oxides for thermoelectric applications}},
  author={Tewari, Abhishek and Dixit, Siddharth and Sahni, Niteesh and Bordas, St{\'e}phane PA},
  journal={Data-Centric Engineering},
  volume={1},
  pages={e8},
  year={2020},
  publisher={Cambridge University Press}
}

@article{zhang2023defect,
  title={{Defect-engineering-stabilized AgSbTe$_2$ with high thermoelectric performance}},
  author={Zhang, Yu and Li, Zhi and Singh, Saurabh and Nozariasbmarz, Amin and Li, Wenjie and Gen{\c{c}}, Aziz and Xia, Yi and Zheng, Luyao and Lee, Seng Huat and Karan, Sumanta Kumar and others},
  journal={Advanced Materials},
  volume={35},
  number={11},
  pages={2208994},
  year={2023},
  publisher={Wiley Online Library}
}

@article{mao2017defect,
  title={{Defect engineering for realizing high thermoelectric performance in n-type Mg$_3$Sb$_2$-based materials}},
  author={Mao, Jun and Wu, Yixuan and Song, Shaowei and Zhu, Qing and Shuai, Jing and Liu, Zihang and Pei, Yanzhong and Ren, Zhifeng},
  journal={ACS Energy Letters},
  volume={2},
  number={10},
  pages={2245--2250},
  year={2017},
  publisher={ACS Publications}
}

@article{xie2022high,
  title={{High Thermoelectric Performance in Chalcopyrite Cu\textsubscript{1-x}Ag\textsubscript{x} GaTe\textsubscript{2}--ZnTe: Nontrivial Band Structure and Dynamic Doping Effect}},
  author={Xie, Hongyao and Liu, Yukun and Zhang, Yinying and Hao, Shiqiang and Li, Zhi and Cheng, Matthew and Cai, Songting and Snyder, G Jeffrey and Wolverton, Christopher and Uher, Ctirad and others},
  journal={Journal of the American Chemical Society},
  volume={144},
  number={20},
  pages={9113--9125},
  year={2022},
  publisher={ACS Publications}
}

@article{pei2017integrating,
  title={{Integrating band structure engineering with all-scale hierarchical structuring for high thermoelectric performance in PbTe system}},
  author={Pei, Yanling and Tan, Gangjian and Feng, Dan and Zheng, Lei and Tan, Qing and Xie, Xiaobing and Gong, Shengkai and Chen, Yue and Li, Jing-Feng and He, Jiaqing and others},
  journal={Advanced Energy Materials},
  volume={7},
  number={3},
  pages={1601450},
  year={2017},
  publisher={Wiley Online Library}
}

@article{zheng2015mechanically,
  title={{Mechanically robust BiSbTe alloys with superior thermoelectric performance: a case study of stable hierarchical nanostructured thermoelectric materials}},
  author={Zheng, Yun and Zhang, Qiang and Su, Xianli and Xie, Hongyao and Shu, Shengcheng and Chen, Tianle and Tan, Gangjian and Yan, Yonggao and Tang, Xinfeng and Uher, Ctirad and others},
  journal={Advanced Energy Materials},
  volume={5},
  number={5},
  pages={1401391},
  year={2015},
  publisher={Wiley Online Library}
}

@article{zhang2024high,
  title={{High wide-temperature-range thermoelectric performance in GeTe through hetero-nanostructuring}},
  author={Zhang, Qingtang and Ying, Pan and Farrukh, Aftab and Gong, Yaru and Liu, Jizi and Huang, Xinqi and Li, Di and Wang, Meiyu and Chen, Guang and Tang, Guodong},
  journal={Acta Materialia},
  volume={276},
  pages={120132},
  year={2024},
  publisher={Elsevier}
}

@article{katsura2025starrydata,
  title={Starrydata: from published plots to shared materials data},
  author={Katsura, Yukari and Kumagai, Masaya and Mato, Tomoya and Takada, Yu and Ando, Yuki and Fujita, Erina and Hosono, Fumikazu and Koyama, Eiji and Mudasar, Farhan and Phuong, Ton Nu Thanh and others},
  journal={Science and Technology of Advanced Materials: Methods},
  volume={5},
  number={1},
  pages={2506976},
  year={2025},
  publisher={Taylor \& Francis}
}

@article{liu2018phase,
  title={{Phase-transition temperature suppression to achieve cubic GeTe and high thermoelectric performance by Bi and Mn codoping}},
  author={Liu, Zihang and Sun, Jifeng and Mao, Jun and Zhu, Hangtian and Ren, Wuyang and Zhou, Jingchao and Wang, Zhiming and Singh, David J and Sui, Jiehe and Chu, Ching-Wu and others},
  journal={Proceedings of the National Academy of Sciences},
  volume={115},
  number={21},
  pages={5332--5337},
  year={2018},
  publisher={National Academy of Sciences}
}

@article{levin2013electronic,
  title={{Electronic and thermal transport in GeTe: A versatile base for thermoelectric materials}},
  author={Levin, EM and Besser, MF and Hanus, Riley},
  journal={Journal of Applied Physics},
  volume={114},
  number={8},
  year={2013},
  publisher={AIP Publishing}
}

@article{li2017simultaneous,
  title={{Simultaneous optimization of carrier concentration and alloy scattering for ultrahigh performance GeTe thermoelectrics}},
  author={Li, Juan and Chen, Zhiwei and Zhang, Xinyue and Yu, Hulei and Wu, Zihua and Xie, Huaqing and Chen, Yue and Pei, Yanzhong},
  journal={Advanced Science},
  volume={4},
  number={12},
  pages={1700341},
  year={2017},
  publisher={Wiley Online Library}
}

@article{wu2017resonant,
  title={{Resonant level-induced high thermoelectric response in indium-doped GeTe}},
  author={Wu, Lihua and Li, Xin and Wang, Shanyu and Zhang, Tiansong and Yang, Jiong and Zhang, Wenqing and Chen, Lidong and Yang, Jihui},
  journal={NPG Asia Materials},
  volume={9},
  number={1},
  pages={e343--e343},
  year={2017},
  publisher={Nature Publishing Group}
}

@article{dou2021lone,
  title={{Lone-pair engineering: Achieving ultralow lattice thermal conductivity and enhanced thermoelectric performance in Al-doped GeTe-based alloys}},
  author={Dou, Y and Li, J and Xie, Y and Wu, X and Hu, L and Liu, F and Ao, W and Liu, Y and Zhang, C},
  journal={Materials Today Physics},
  volume={20},
  pages={100497},
  year={2021},
  publisher={Elsevier}
}

@article{hong2018realizing,
  title={{Realizing zT of 2.3 in Ge$_{1-x-y}$Sb$_x$In$_y$Te via reducing the phase-transition temperature and introducing resonant energy doping}},
  author={Hong, Min and Chen, Zhi-Gang and Yang, Lei and Zou, Yi-Chao and Dargusch, Matthew S and Wang, Hao and Zou, Jin},
  journal={Advanced materials},
  volume={30},
  number={11},
  pages={1705942},
  year={2018},
  publisher={Wiley Online Library}
}

@article{lei2019ultrafast,
  title={{Ultrafast synthesis of Te-doped CoSb$_3$ with excellent thermoelectric properties}},
  author={Lei, Ying and Gao, Wensheng and Zheng, Rui and Li, Yu and Chen, Wen and Zhang, Libo and Wan, Rundong and Zhou, Hongwei and Liu, Zhiyuan and Chu, Paul K},
  journal={ACS Applied Energy Materials},
  volume={2},
  number={6},
  pages={4477--4485},
  year={2019},
  publisher={ACS Publications}
}

@article{shi2011multiple,
  title={Multiple-filled skutterudites: high thermoelectric figure of merit through separately optimizing electrical and thermal transports},
  author={Shi, Xun and Yang, Jiong and Salvador, James R and Chi, Miaofang and Cho, Jung Y and Wang, Hsin and Bai, Shengqiang and Yang, Jihui and Zhang, Wenqing and Chen, Lidong},
  journal={Journal of the American Chemical Society},
  volume={133},
  number={20},
  pages={7837--7846},
  year={2011},
  publisher={ACS Publications}
}

@article{TbCuTe2,
  title={{Syntheses, structures, and thermoelectric properties of ternary tellurides: RECuTe$_2$ (RE=Tb--Er)}},
  author={Lin, Hua and Chen, Hong and Ma, Ni and Zheng, Yu-Jun and Shen, Jin-Ni and Yu, Ju-Song and Wu, Xin-Tao and Wu, Li-Ming},
  journal={Inorganic Chemistry Frontiers},
  volume={4},
  number={8},
  pages={1273--1280},
  year={2017},
  publisher={Royal Society of Chemistry}
}

@article{AgBiS2,
  title={{Thermoelectric properties of AgBiS$_2$: Unveiling the origin of ultra-low lattice thermal conductivity and optimization strategies for electrical performance}},
  author={An, Hongxu and Wang, Dongyang and He, Wenke},
  journal={Applied Physics Letters},
  volume={127},
  number={14},
  year={2025},
  publisher={AIP Publishing}
}

@article{In4SnSe4,
  title={{A promising thermoelectrics In$_4$SnSe$_4$ with a wide bandgap and cubic structure composited by layered SnSe and In$_4$Se$_3$}},
  author={Shi, Haonan and Guo, Changrong and Qin, Bingchao and Wang, Guangtao and Wang, Dongyang and Zhao, Li-Dong},
  journal={Journal of Materiomics},
  volume={8},
  number={5},
  pages={982--991},
  year={2022},
  publisher={Elsevier}
}

@article{Sr3AlSb3,
  title={{Thermoelectric Properties and Electronic Structure of the Zintl-Phase Sr$_3$AlSb$_3$}},
  author={Zevalkink, Alex and Pomrehn, Gregory and Takagiwa, Yoshiki and Swallow, Jessica and Snyder, G Jeffrey},
  journal={ChemSusChem},
  volume={6},
  number={12},
  pages={2316--2321},
  year={2013},
  publisher={Wiley Online Library}
}

@article{CuSbS2,
  title={{Synthesis, structure, Te alloying, and physical properties of CuSbS$_2$}},
  author={Hobbis, Dean and Wei, Kaya and Wang, Hsin and Martin, Joshua and Nolas, George S},
  journal={Inorganic Chemistry},
  volume={56},
  number={22},
  pages={14040--14044},
  year={2017},
  publisher={ACS Publications}
}

@article{caillat1996properties,
  title={{Properties of single crystalline semiconducting CoSb$_3$}},
  author={Caillat, T and Borshchevsky, A and Fleurial, J-P},
  journal={Journal of Applied Physics},
  volume={80},
  number={8},
  pages={4442--4449},
  year={1996},
  publisher={American Institute of Physics}
}

@article{park2014thermoelectric,
  title={{Thermoelectric properties of Yb-filled CoSb$_3$ skutterudites}},
  author={Park, Kwan-Ho and Seo, Won-Seon and Shin, Dong-Kil and Kim, Il-Ho},
  journal={Journal of the Korean Physical Society},
  volume={65},
  number={4},
  pages={491--495},
  year={2014},
  publisher={Springer}
}

@article{sun2016effect,
  title={{Effect of high-temperature and high-pressure processing on the structure and thermoelectric properties of clathrate Ba$_8$Ga$_{16}$Ge$_{30}$}},
  author={Sun, Bing and Jia, Xiaopeng and Huo, Dexuan and Sun, Hairui and Zhang, Yuewen and Liu, Binwu and Liu, Haiqiang and Kong, Lingjiao and Liu, Baomin and Ma, Hongan},
  journal={The Journal of Physical Chemistry C},
  volume={120},
  number={18},
  pages={10104--10110},
  year={2016},
  publisher={ACS Publications}
}

@article{harnwunggmoung2010high,
  title={High-temperature thermoelectric properties of thallium-filled skutterudites},
  author={Harnwunggmoung, Adul and Kurosaki, Ken and Muta, Hiroaki and Yamanaka, Shinsuke},
  journal={Applied Physics Letters},
  volume={96},
  number={20},
  year={2010},
  publisher={AIP Publishing}
}

@article{kim2019enhanced,
  title={{Enhanced thermoelectric properties of Bi$_{0.5}$Sb$_{1.5}$Te$_3$ composites with in-situ formed senarmontite Sb$_2$O$_3$ nanophase}},
  author={Kim, Eun Bin and Dharmaiah, Peyala and Lee, Kap-Ho and Lee, Chul-Hee and Lee, Jong-Hyeon and Yang, Jae-Kyo and Jang, Dae-Hwan and Kim, Dong-Soo and Hong, Soon-Jik},
  journal={Journal of Alloys and Compounds},
  volume={777},
  pages={703--711},
  year={2019},
  publisher={Elsevier}
}

@article{tanusilp2020beneficial,
  title={{Beneficial influence of iodine substitution on the thermoelectric properties of Mo$_3$Sb$_7$}},
  author={Tanusilp, Sora-at and Wongprakarn, Suphagrid and Kidkhunthod, Pinit and Ohishi, Yuji and Muta, Hiroaki and Kurosaki, Ken},
  journal={Journal of Applied Physics},
  volume={127},
  number={10},
  year={2020},
  publisher={AIP Publishing}
}

@article{xu2021substantial,
  title={{Substantial thermoelectric enhancement achieved by manipulating the band structure and dislocations in Ag and La co-doped SnTe}},
  author={Xu, Wenjing and Zhang, Zhongwei and Liu, Chengyan and Gao, Jie and Ye, Zhenyuan and Chen, Chunguang and Peng, Ying and Bai, Xiaobo and Miao, Lei},
  journal={Journal of Advanced Ceramics},
  volume={10},
  number={4},
  pages={860--870},
  year={2021},
  publisher={Springer}
}

@article{dong2010microwave,
  title={{Microwave-assisted rapid synthesis of Sb$_2$Te$_3$ nanosheets and thermoelectric properties of bulk samples prepared by spark plasma sintering}},
  author={Dong, Guo-Hui and Zhu, Ying-Jie and Chen, Li-Dong},
  journal={Journal of Materials Chemistry},
  volume={20},
  number={10},
  pages={1976--1981},
  year={2010},
  publisher={Royal Society of Chemistry}
}

@article{yang2015facile,
  title={{A facile surfactant-assisted reflux method for the synthesis of single-crystalline Sb$_2$Te$_3$ nanostructures with enhanced thermoelectric performance}},
  author={Yang, Heng Quan and Miao, Lei and Liu, Cheng Yan and Li, Chao and Honda, Sawao and Iwamoto, Yuji and Huang, Rong and Tanemura, Sakae},
  journal={ACS Applied Materials \& Interfaces},
  volume={7},
  number={26},
  pages={14263--14271},
  year={2015},
  publisher={ACS Publications}
}

@article{hu2015enhanced,
  title={{Enhanced figure of merit in antimony telluride thermoelectric materials by In--Ag co-alloying for mid-temperature power generation}},
  author={Hu, LP and Zhu, TJ and Yue, XQ and Liu, XH and Wang, YG and Xu, ZJ and Zhao, XB},
  journal={Acta Materialia},
  volume={85},
  pages={270--278},
  year={2015},
  publisher={Elsevier}
}

@article{saleemi2014evaluation,
  title={{Evaluation of the structure and transport properties of nanostructured antimony telluride (Sb$_2$Te$_3$)}},
  author={Saleemi, Mohsin and Ruditskiy, A and Toprak, MS and Stingaciu, Marian and Johnsson, Mats and Kretzschmar, I and Jacquot, A and J{\"a}gle, M and Muhammed, Mamoun},
  journal={Journal of Electronic Materials},
  volume={43},
  number={6},
  pages={1927--1932},
  year={2014},
  publisher={Springer}
}

@article{nolas1999skutterudites,
  title={Skutterudites: A phonon-glass-electron crystal approach to advanced thermoelectric energy conversion applications},
  author={Nolas, GS and Morelli, DT and Tritt, Terry M},
  journal={Annual Review of Materials Science},
  volume={29},
  number={1},
  pages={89--116},
  year={1999},
  publisher={Annual Reviews 4139 El Camino Way, PO Box 10139, Palo Alto, CA 94303-0139, USA}
}

@article{takabatake2014phonon,
  title={Phonon-glass electron-crystal thermoelectric clathrates: Experiments and theory},
  author={Takabatake, Toshiro and Suekuni, Koichiro and Nakayama, Tsuneyoshi and Kaneshita, Eiji},
  journal={Reviews of Modern Physics},
  volume={86},
  number={2},
  pages={669--716},
  year={2014},
  publisher={APS}
}

@article{snyder2004disordered,
  title={{Disordered zinc in Zn$_4$Sb$_3$ with phonon-glass and electron-crystal thermoelectric properties}},
  author={Snyder, G Jeffrey and Christensen, Mogens and Nishibori, Eiji and Caillat, Thierry and Iversen, Bo Brummerstedt},
  journal={Nature Materials},
  volume={3},
  number={7},
  pages={458--463},
  year={2004},
  publisher={Nature Publishing Group UK London}
}

@article{liu2012copper,
  title={Copper ion liquid-like thermoelectrics},
  author={Liu, Huili and Shi, Xun and Xu, Fangfang and Zhang, Linlin and Zhang, Wenqing and Chen, Lidong and Li, Qiang and Uher, Ctirad and Day, Tristan and Snyder, G Jeffrey},
  journal={Nature materials},
  volume={11},
  number={5},
  pages={422--425},
  year={2012},
  publisher={Nature Publishing Group UK London}
}

@article{wang2011enhanced,
  title={{Enhanced performances of melt spun Bi$_2$(Te,Se)$_3$ for n-type thermoelectric legs}},
  author={Wang, Shanyu and Xie, Wenjie and Li, Han and Tang, Xinfeng},
  journal={Intermetallics},
  volume={19},
  number={7},
  pages={1024--1031},
  year={2011},
  publisher={Elsevier}
}

@article{xue2020off,
  title={{Off-stoichiometry effects on the thermoelectric properties of Cu\textsubscript{2+$\delta$}Se (-0.1$\leq\delta\leq$0.05) compounds synthesized by a high-pressure and high-temperature method}},
  author={Xue, Lisha and Shen, Weixia and Zhang, Zhuangfei and Shen, Manjie and Ji, Wenting and Fang, Chao and Zhang, Yuewen and Jia, Xiaopeng},
  journal={CrystEngComm},
  volume={22},
  number={4},
  pages={695--700},
  year={2020},
  publisher={Royal Society of Chemistry}
}

@article{dong2019medium,
  title={{Medium-temperature thermoelectric GeTe: vacancy suppression and band structure engineering leading to high performance}},
  author={Dong, Jinfeng and Sun, Fu-Hua and Tang, Huaichao and Pei, Jun and Zhuang, Hua-Lu and Hu, Hai-Hua and Zhang, Bo-Ping and Pan, Yu and Li, Jing-Feng},
  journal={Energy \& Environmental Science},
  volume={12},
  number={4},
  pages={1396--1403},
  year={2019},
  publisher={Royal Society of Chemistry}
}

@article{falkenbach2015tin,
  title={{Tin Telluride-Based Nanocomposites of the Type AgSn$_m$BiTe$_{2+m}$ (BTST-m) as Effective Lead-Free Thermoelectric Materials}},
  author={Falkenbach, Oliver and Schmitz, Andreas and Dankwort, Torben and Koch, Guenter and Kienle, Lorenz and Mueller, Eckhard and Schlecht, Sabine},
  journal={Chemistry of Materials},
  volume={27},
  number={21},
  pages={7296--7305},
  year={2015},
  publisher={ACS Publications}
}

@article{el2021unraveling,
  title={{Unraveling the thermoelectric performance of Bismuth Antimony/graphene nanocomposite synthesized by spark plasma extrusion}},
  author={El-Asfoury, Mohamed S and Abdou, Shaban M and Nassef, Ahmed},
  journal={Journal of Alloys and Compounds},
  volume={887},
  pages={161399},
  year={2021},
  publisher={Elsevier}
}

@article{liang2021ductile,
  title={{Ductile inorganic amorphous/crystalline composite Ag$_4$TeS with phonon-glass electron-crystal transport behavior and excellent stability of high thermoelectric performance on plastic deformation}},
  author={Liang, Xin and Chen, Chuang},
  journal={Acta Materialia},
  volume={218},
  pages={117231},
  year={2021},
  publisher={Elsevier}
}

\end{document}